# Exoplanet Biosignatures:
# A Review of Remotely Detectable Signs of Life

Edward W. Schwieterman,[1–5] Nancy Y. Kiang,[3,6] Mary N. Parenteau,[3,7] Chester E. Harman,[3,6,8]
Shiladitya DasSarma,[9,10] Theresa M. Fisher,[11] Giada N. Arney,[3,12] Hilairy E. Hartnett,[11,13]
Christopher T. Reinhard,[4,14] Stephanie L. Olson,[1,4] Victoria S. Meadows,[3,15]
Charles S. Cockell,[16,17] Sara I. Walker,[5,11,18,19] John Lee Grenfell,[20]
Siddharth Hegde,[21,22] Sarah Rugheimer,[23] Renyu Hu,[24,25] and Timothy W. Lyons[1,4]

## Abstract

In the coming years and decades, advanced space- and ground-based observatories will allow an unprecedented opportunity to probe the atmospheres and surfaces of potentially habitable exoplanets for signatures of life. Life on Earth, through its gaseous products and reflectance and scattering properties, has left its fingerprint on the spectrum of our planet. Aided by the universality of the laws of physics and chemistry, we turn to Earth's biosphere, both in the present and through geologic time, for analog signatures that will aid in the search for life elsewhere. Considering the insights gained from modern and ancient Earth, and the broader array of hypothetical exoplanet possibilities, we have compiled a comprehensive overview of our current understanding of potential exoplanet biosignatures, including gaseous, surface, and temporal biosignatures. We additionally survey biogenic spectral features that are well known in the specialist literature but have not yet been robustly vetted in the context of exoplanet biosignatures. We briefly review advances in assessing biosignature plausibility, including novel methods for determining chemical disequilibrium from remotely obtainable data and assessment tools for determining the minimum biomass required to maintain short-lived biogenic gases as atmospheric signatures. We focus particularly on advances made since the seminal review by Des Marais *et al.* The purpose of this work is not to propose new biosignature strategies, a goal left to companion articles in this series, but to review the current literature, draw meaningful connections between seemingly disparate areas, and clear the way for a path forward. Key Words: Exoplanets—Biosignatures—Habitability markers—Photosynthesis—Planetary surfaces—Atmospheres—Spectroscopy—Cryptic biospheres—False positives. Astrobiology 18, 663–708.

[1]Department of Earth Sciences, University of California, Riverside, California.
[2]NASA Postdoctoral Program, Universities Space Research Association, Columbia, Maryland.
[3]NASA Astrobiology Institute, Virtual Planetary Laboratory Team, Seattle, Washington.
[4]NASA Astrobiology Institute, Alternative Earths Team, Riverside, California.
[5]Blue Marble Space Institute of Science, Seattle, Washington.
[6]NASA Goddard Institute for Space Studies, New York, New York.
[7]NASA Ames Research Center, Exobiology Branch, Mountain View, California.
[8]Department of Applied Physics and Applied Mathematics, Columbia University, New York, New York.
[9]Department of Microbiology and Immunology, University of Maryland School of Medicine, Baltimore, Maryland.
[10]Institute of Marine and Environmental Technology, University System of Maryland, Baltimore, Maryland.
[11]School of Earth and Space Exploration, Arizona State University, Tempe, Arizona.
[12]Planetary Systems Laboratory, NASA Goddard Space Flight Center, Greenbelt, Maryland.
[13]School of Molecular Sciences, Arizona State University, Tempe, Arizona.
[14]School of Earth and Atmospheric Sciences, Georgia Institute of Technology, Atlanta, Georgia.
[15]Astronomy Department, University of Washington, Seattle, Washington.
[16]University of Edinburgh School of Physics and Astronomy, Edinburgh, United Kingdom.
[17]UK Centre for Astrobiology, Edinburgh, United Kingdom.
[18]Beyond Center for Fundamental Concepts in Science, Arizona State University, Tempe, Arizona.
[19]ASU-Santa Fe Institute Center for Biosocial Complex Systems, Arizona State University, Tempe, Arizona.
[20]Institut für Planetenforschung (PF), Deutsches Zentrum für Luft und Raumfahrt (DLR), Berlin, Germany.
[21]Carl Sagan Institute, Cornell University, Ithaca, New York.
[22]Cornell Center for Astrophysics and Planetary Science, Cornell University, Ithaca, New York.
[23]Department of Earth and Environmental Sciences, University of St. Andrews, St. Andrews, United Kingdom.
[24]Jet Propulsion Laboratory, California Institute of Technology, Pasadena, California.
[25]Division of Geological and Planetary Sciences, California Institute of Technology, Pasadena, California.

---

**Table of Contents**

## 1. Introduction

THE SEARCH FOR life beyond the Solar System is a significant motivator for the detection and characterization of extrasolar planets around nearby stars. We are poised at the transition between exoplanet detection and demographic studies and the detailed characterization of exoplanet atmospheres and surfaces. Transit and radial velocity surveys have confirmed the existence of thousands of exoplanets (Akeson *et al.*, 2013; Batalha, 2014; Morton *et al.*, 2016) with well over a dozen located within the circumstellar habitable zones (HZs) of their host stars (*e.g.*, Kane *et al.*, 2016). Planets with masses and radii consistent with rocky compositions and likely to contain secondary, volcanically outgassed atmospheres have been found in nearby stellar systems (Berta-Thompson *et al.*, 2015; Wright *et al.*, 2016), some of which reside in the HZ of their host star such as Proxima Centauri b (Anglada-Escudé *et al.*, 2016); TRAPPIST-1 e, f, and g (Gillon *et al.*, 2017); and LHS 1140b (Dittmann *et al.*, 2017).

Those planets that transit their stars are excellent candidates for atmospheric characterization through transmission spectroscopy with the upcoming James Webb Space Telescope (JWST) set to launch in 2020 (Deming *et al.*, 2009; Stevenson *et al.*, 2016). Planets with sufficient planet–star separations will likewise be excellent targets for direct-imaging spectroscopy. Space-based telescope missions with the capability of measuring directly imaged spectra of potentially habitable exoplanets are in their science-definition stages (*e.g.*, Dalcanton *et al.*, 2015; Mennesson *et al.*, 2016). Ground-based observers are also devising instrumentation and techniques for current and future observatories that will have the capacity to image Earth-sized planets around nearby stars (Kawahara *et al.*, 2012; Snellen *et al.*, 2013, 2015; Lovis *et al.*, 2016).

The characterization of exoplanetary atmospheres and surfaces in search of remotely detectable biosignatures is a mandate of the NASA Astrobiology Program (Des Marais *et al.*, 2008; Hays *et al.*, 2015; Voytek, 2016). In support of this mandate, and the future observations and missions that will fulfill it, we have compiled an updated review of exoplanet biosignatures. In anticipation of the planned (but later canceled) Terrestrial Planet Finder (TPF) mission, Des Marais *et al.* (2002) gave us one of the most comprehensive reviews now available. Our review will emphasize advances in exoplanet biosignature science since the Des Marais *et al.* (2002) review. These advances have taken many forms, from those demonstrating the detectability of Earth's own biosphere using updated spectral models and data–model comparisons (*e.g.*, Robinson *et al.*, 2011), to the application of photochemical models that test expected changes in concentration of biogenic gases for an Earth-like biosphere around stars with vastly different properties than our own (*e.g.*, Segura *et al.*, 2005). Other advances include evaluations of spectral signatures in the context of plausible biomasses (*e.g.*, Seager *et al.*, 2013a) and metrics for chemical disequilibrium (*e.g.*, Krissansen-Totton *et al.*, 2016a).

It is beyond the scope of this contribution to recapitulate every detail of the aforementioned studies; instead we provide a starting point that exposes readers to general concepts developed in past work. We additionally draw new connections between existing studies and bring forward relevant specialist literature that has not yet been examined in the context of exoplanet biosignatures. Recommendations for future directions are left to the companion articles in this series. This review consists of the following sections: (1) requirements for life, biosignature definitions, and biosignature categories; (2) evaluating planetary habitability; (3) an overview of terrestrial exoplanet modeling studies; (4) gaseous biosignatures, including descriptions of specific gases; (5) surface biosignatures, including description of specific potential reflectance signatures; (6) temporal biosignatures; (7) methods of assessing biosignature plausibility; (8) cryptic biospheres and "false negatives" for life; (9) prospects for detecting exoplanet biosignatures; and (10) a summary of this review.

### 1.1. Requirements for life

The search for life beyond Earth is one of the most monumental and consequential endeavors on which humanity has ever embarked. It is also a search that is fraught with intricacies and complexities. Our definition of life is necessarily limited by our understanding of life on Earth; however, we are aided by the universality of the laws of physics and chemistry. Through this notion of universality, a consensus has emerged that life requires three essential components: (1) an energy source to drive metabolic reactions, (2) a liquid solvent to mediate these reactions, and (3) a suite of nutrients both to build biomass and to produce enzymes that catalyze metabolic reactions (see Cockell *et al.*, 2016, for an expanded discussion of these requirements).

The study of life on Earth and the general principles of chemistry and physics further suggest, although less strongly, that the liquid solvent is likely to be water, both because of its cosmic abundance (it is one of the most cosmically abundance molecules, consisting of the first [H] and third [O] most abundant elements) and its distinct physicochemical properties that make it highly suitable for mediating macromolecular interactions. While one of water's essential properties is its oft-cited ability to act as a solvent for polar molecules, promoted by its unique ability to engage in hydrogen bonding, water has much more expansive, active and, at times, subtle, roles within known living processes. For example, water plays an essential role in protein folding, protein substrate binding, enzyme actions, the rapid transport of protons in aqueous solution, maintaining the structural stability of proteins and DNA/RNA, and the inhomogeneous segregation of salt ions at cellular interfaces (for a more in-depth discussion, see Ball, 2008, 2013). Carbon chemistry is likewise favored as a basis for biomass because carbon has a high cosmic abundance and carries the ability to form an inordinate number of complex molecules. These last two assumptions are made here provisionally, with the acknowledgment that while alternative biochemistries may exist, their plausibility has not yet been convincingly demonstrated (nor their potential biosignatures explicated). Further constraints on the development and persistence of life likely exist, although they are less precisely enumerable.

The most defining aspect of life is the capacity for evolution, which is necessary to adapt organisms to changing environmental conditions. This requirement to engage in evolutionary adaptation likely requires complex functional molecules that preclude environments too extreme to allow the formation and long-term persistence of such molecules, although this environmental space is yet to be completely circumscribed (see discussion in Des Marais, 2013). The unsettled nature of the definition of life is beyond the scope of this article, and for an expanded discussion of this topic, the reader is referred to Walker *et al.* (2018, this issue), who provide a more extensive presentation on the definition of life and propose new and diverse conceptual frameworks for expanding the search for life in the Universe.

### 1.2. Exoplanet biosignature definitions

A biosignature is nominally defined as an ''object, substance, and/or pattern whose origin specifically requires a biological agent'' (Des Marais and Walter, 1999; Des Marais *et al.*, 2008). A sign of life from an exoplanet may manifest itself as a spectroscopic signal (or signals), a measurement that will have a stated uncertainty and potentially a range of explanations (including measurement error). That signal may be used to infer the presence of a gas or surface feature, which then may be interpreted as originating from a living process. As a matter of definition, we may ask whether the biosignature is the measured spectral signature or the inferred presence of the gas based on that signature. Or, rather, is the biosignature a further inference that a living process must have been involved in the production of the gas or surface feature, perhaps through the collection of additional remotely sensed information? If latter, what level of certainty is required to designate the feature(s) a ''biosignature?'' In other words, can something be considered a biosignature if there is a nonzero probability that it is not produced by life?

Use of the term ''biosignature'' in the context of astronomical observations varies widely, but it is mostly commonly understood to be the presence of a gas or other feature that is indicative of a biological agent. In this work, we relax the formal definition of ''biosignature'' with the understanding that in practice, for almost any conceivable circumstance, a prospective exoplanet biosignature will always be a *potential* biosignature with other possible explanations (unless technological, but see below). To state this another way, a gas may be a ''biosignature gas,'' even if the gas may have nonbiological sources. Our challenge then would be to test and ideally prove a biological origin. An alternative position would be that there can be *no* exoplanet biosignatures, since *all* hypothetical biosignatures could have false positives (nonbiological origins), and ground truth verification of the biogenicity of any remotely detected signature will be unachievable for the determinable future.

The admission that all proposed exoplanet biosignatures are potential biosignatures in current practice is necessary and inescapable, and protects against false confidence when the full range of abiotic chemistries that may produce false positives is unknown. It is further supported by the experience of researchers studying microfossil evidence for Earth's earliest life-forms, which can often be inconclusive or open to a range of interpretations (*e.g.*, Schopf, 1993; Garcia-Ruiz *et al.*, 2003; Brasier *et al.*, 2015), but none-theless providing useful information about the evolution of life on our own planet. At the same time, rejection of the potential biosignature concept, as a philosophical choice, may also be unnecessarily pessimistic. Our preference is to find the right balance—that is, to define possibilities and to constrain them with every tool available toward identifying biology as the origin of a putative biosignature, while keeping our minds open to the possibility that such vetting down the road may be beyond technology or scientific understanding available at the time of first observation. An escape from this dilemma would be provided by radio or nonradio "technosignatures"—unambiguous signs of technological civilization explored by practitioners of the Search for Extraterrestrial Intelligence (SETI; see, *e.g.*, Tarter, 2001; Cabrol, 2016). While an important and compelling area of study, SETI and technosignatures are beyond the scope of this review, which focuses on signatures of nontechnological life. Here we use the term "biosignature" to refer to nontechnological signs of life unless otherwise noted.

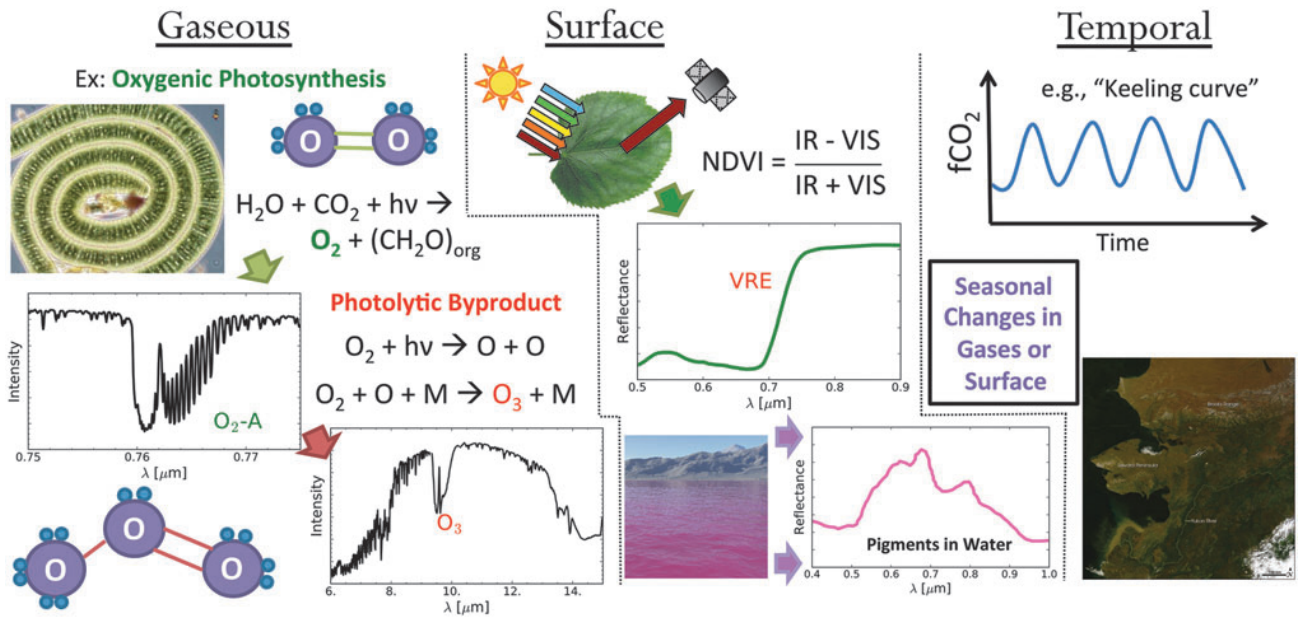

**FIG. 1.** Summary of gaseous, surface, and temporal biosignatures. Left panel: gaseous biosignatures are direct or indirect products of biological processes. One example is molecular $O_2$ generated as a by-product of photosynthesis that is then photochemically processed into $O_3$ in the stratosphere. Middle panel: surface biosignatures are the spectral signatures imparted by reflected light that interacts directly with living material. One example is the well-known VRE produced by plants and the associated NDVI used for mapping surface vegetation on Earth (Tucker, 1979). Right panel: time-dependent changes in observable quantities, including gas concentrations or surface albedo features, may represent a temporal biosignature if they can be linked to the response of a biosphere to a seasonal or diurnal change. An example is the seasonal oscillation of $CO_2$ as a response to the seasonal growth and decay of vegetation (*e.g.*, Keeling *et al.*, 1976). This figure is reproduced with permission from Schwieterman (2016). Subimage credits: NASA and the Encyclopedia of Life (EOL). NDVI, Normalized Difference Vegetation Index; $O_2$, oxygen; $O_3$, ozone; VRE, vegetation red edge.

### 1.3. Biosignature categories

There is currently no universally accepted scheme for classifying the vast array of potential exoplanet biosignatures. For convenience, we group biosignatures into three broad categories following a suggestion by Meadows (2006, 2008): gaseous, surface, and temporal biosignatures (Fig. 1). In this scheme, gaseous biosignatures are direct or indirect products of metabolism, surface biosignatures are spectral features imparted on radiation reflected or scattered by organisms, and temporal biosignatures are modulations in measurable quantities that can be linked to the actions and time-dependent patterns of a biosphere. Gaseous, surface, and temporal biosignatures are reviewed in depth in Sections 4, 5, and 6, respectively. Frameworks for further classifying gaseous signatures are reviewed and proposed in a companion article (Walker *et al.*, 2018).

### 2. Evaluating Planetary Habitability

The focus of this work is on summarizing proposed exoplanet biosignatures rather than on definitions and metrics for habitability. A full discussion of habitability would require reviewing the complex interplay among instellation, atmospheric dynamics, greenhouse gases, planetary tectonics, orbital stability, ice-albedo feedbacks, the remote detectability of these processes, and many other topics beyond the scope of this review. On the other hand, evaluation of biosignatures must include some discussion of habitability, both because inhabited planets are habitable and because detectable metrics for habitability will assist in the interpretation of potential biosignatures. We adopt the definition that habitable planets are those capable of supporting stable liquid water on the surface. There may be a wide variety of atmospheric compositions that can achieve this result, including $N_2$-, $CO_2$-, and $H_2$-dominated atmospheres (Kasting *et al.*, 1993; Pierrehumbert and Gaidos, 2011; Kopparapu *et al.*, 2013; Seager, 2013; Ramirez and Kaltenegger, 2017).

Evaluation of potential habitability is assisted by the concept of the Habitable Zone (HZ), defined as the range of distances, or annulus, around a star that would allow a planet with a given atmosphere to maintain surface liquid water (Fig. 2). This definition allows for a rapid assessment of potential habitability if observables such as semimajor axis and stellar luminosity can be adequately constrained. (Stellar luminosity may be directly measured or estimated from other measured stellar parameters such as effective temperature and radius, although there are notable large errors for the latter that may be propagated into HZ estimates, see Kane, 2014.) The most common definition of the HZ assumes an $N_2$-$CO_2$-$H_2O$ atmosphere with a carbonate–silicate feedback cycle (Walker *et al.*, 1981; Kasting *et al.*, 1993; Kopparapu *et al.*, 2013) that acts as a planetary thermostat, as it is believed to do on Earth. In this conception of the HZ, planetary temperature is primarily controlled by greenhouse absorption via $CO_2$ and $H_2O$, and the overall planetary albedo—a product of atmospheric mass and composition, cloud cover and composition, surface albedo, and stellar temperature.

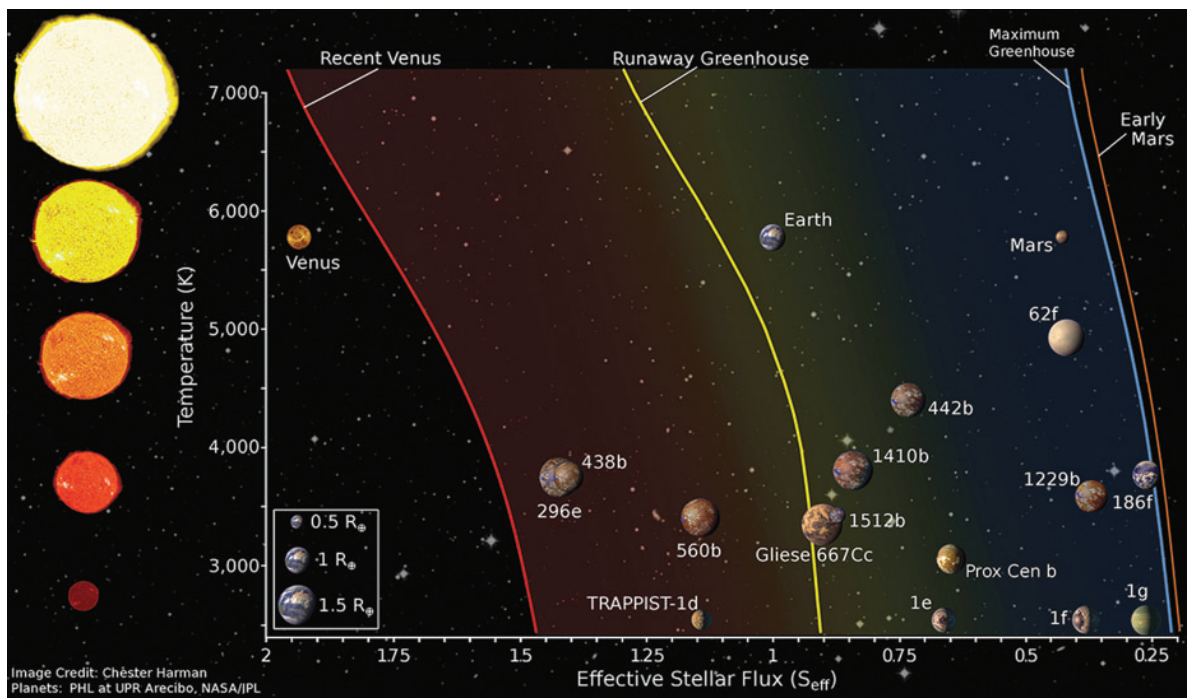

**FIG. 2.** The circumstellar habitable zone. Planets within the Habitable Zone have the capacity to maintain stable surface liquid water assuming an $N_2$-$CO_2$-$H_2O$ atmosphere and a carbonate–silicate feedback cycle (*e.g.*, Kasting *et al.*, 1993; Kopparapu *et al.*, 2013). The assumption of surface liquid water is important because it suggests a biosphere, if present, would be in direct contact with the atmosphere, allowing the buildup of potentially detectable biosignatures in the atmosphere and/or on the surface.

The boundaries of the HZ in terms of stellar instellation will vary as a function of stellar type as the spectral energy distribution of stars of different temperatures will produce different effective planetary albedos even for a planetary atmosphere of constant composition (*e.g.*, less or more blue light to Rayleigh scatter, and radiation shifted into or out of primary gas absorption bands). It is important to note that even this definition of the HZ depends on factors such as planetary gravity and atmospheric mass, which can alter the greenhouse effect due to pressure broadening effects (Kopparapu *et al.*, 2014), and ice cover and surface composition (Shields *et al.*, 2013, 2014). Other definitions of the HZ are much broader (*e.g.*, Seager, 2013) and include $H_2$-dominated atmospheres where $H_2$-$H_2$ collisionally induced absorption greatly extends the outer edge of the HZ (Pierrehumbert and Gaidos, 2011), possibly to interstellar space (Stevenson, 1999), and dry atmospheres that press the inner edge of the HZ closer to the star (Abe *et al.*, 2011; Zsom *et al.*, 2013). Until recently, most assessments of the HZ have been made with relatively simple one-dimensional (1D) radiative–convective models. Newer work using more advanced three-dimensional (3D) general circulation models (GCMs), however, suggests more optimistic limits at the inner edge of the HZ (Yang *et al.*, 2013, 2014; Leconte *et al.*, 2013a, 2013b; Kopparapu *et al.*, 2016; Shields *et al.*, 2016b) while also showing perhaps more pessimistic results for the outer edge compared with 1D results (Wolf, 2017).

An additional challenge at the outer edge of the HZ is presented by "limit cycles"—oscillations between globally glaciated and climatically warm states resulting from the balance of warming from $CO_2$ outgassing and cooling from $CO_2$ subduction over the carbonate–silicate cycle and consequent changes in albedo from planetary glaciation and deglaciation (Tajika, 2007; Kadoya and Tajika, 2014; Menou, 2015; Haqq-Misra *et al.*, 2016; Paradise and Menou, 2017). Limit cycles have been investigated by a hierarchy of climate models, including simple energy balance models, 1D radiative–convective models, and 3D GCMs. Transiently habitable states at the outer edge of the HZ due to limit cycles may preclude complex or even simple life depending on the duration of warm and cool states. The occurrence of limit cycles will depend on planetary parameters such as the $CO_2$ outgassing rate, the incident stellar flux, and the spectral energy distribution of the host star, with planets orbiting F stars most susceptible to them (Haqq-Misra *et al.*, 2016). For the purposes of target selection for biosignature searches, conservative definitions of the HZ may be preferred to maximize the probability of success (Kasting *et al.*, 2013), and thus, a preference for targets within the most restrictive HZ limits of 1D and 3D modeling results could be adopted, including consideration of limit cycles. In any case, a habitable planet must at minimum possess liquid water and one (or more) noncondensable greenhouse gases sufficient to warm the surface. The presence of a planet within the HZ is a necessary, but not sufficient, condition for habitability by this definition.

Host star type (or effective temperature) must also be considered when evaluating the potential habitability of planets. While radiative–convective or more advanced GCMs may suggest that a given insolation is appropriate for maintaining surface liquid water, other factors that impact planetary habitability are influenced strongly by stellar mass. The most common consideration in this realm is stellar lifetime, with the common assumption that remotely detectable Earth-like biospheres require hundreds of millions to billions of years to develop. If the stellar lifetime is shorter than this time frame, few if any planets orbiting those stars will have had the requisite time to develop biosignatures. This requirement excludes main sequence stars more luminous than spectral type F or a stellar mass of $\sim 1.4\,M_{sol}$.

At the low end of the mass range ($0.075$–$0.5\,M_{sol}$), M stars represent the most common and long-lived type of star in the galaxy, but also possess properties that pose obstacles to habitability. These include the deleterious impacts of the ultraviolet (UV) and particle (flare) activity of these stars (Segura *et al.*, 2003; Lammer *et al.*, 2007; Davenport *et al.*, 2016; Ribas *et al.*, 2016; Airapetian *et al.*, 2017), their premain sequence evolution (Ramirez and Kaltenegger, 2014; Luger and Barnes, 2015; Tian, 2015), and the impact of tidal heating on planetary climate (Barnes *et al.*, 2009; Driscoll and Barnes, 2015; Bolmont and Mathis, 2016; Bolmont *et al.*, 2017). In addition, the low quiescent (nonflaring) near-ultraviolet (NUV) spectrum of M dwarfs may also drastically limit the rate of prebiotic photoprocesses, creating an obstacle for the origin of life on these worlds (Ranjan *et al.*, 2017). Despite these concerns, a provisional consensus holds that M dwarf stars may indeed possess potentially habitable planets (Tarter *et al.*, 2007; Scalo *et al.*, 2007; see Shields *et al.*, 2016a, for a recent, thorough review of the habitability of planets around M stars). Here we consider planets orbiting within the HZ of FGKM stars as potential targets for habitability assessment and biosignature searches.

Confirming habitability requires further investigation beyond simply determining whether a planet lies within the HZ of its star. The most straightforward determination of planetary habitability would be direct detection of surface liquid water, possibly through the observation of glint if the planet can be observed at large phase angles (*e.g.*, Williams and Gaidos, 2008; Robinson *et al.*, 2010, Zugger *et al.*, 2010; Robinson *et al.*, 2014, but see also Cowan *et al.*, 2012). Ocean–land heterogeneity and rotation rate could also be detected through time-dependent spectrophotometric analysis (Ford *et al.*, 2001; Cowan *et al.*, 2009; Kawahara and Fujii, 2010, 2011; Fujii *et al.*, 2011; Cowan and Strait, 2013), which may also provide indirect evidence for other consequences of continentality, including terrestrial habitats, plate tectonics, and attendant nutrient cycling. Alternatively, the stability of liquid water could be determined indirectly by constraining planetary temperature through midinfrared (MIR) observations (*e.g.*, Robinson *et al.*, 2011) and pressure by retrievals based on the Rayleigh scattering slope (Benneke and Seager, 2012) or through highly density-dependent collisional or dimer absorption features of primary atmospheric constituents such as N$_2$ or oxygen (O$_2$) (Pallé *et al.*, 2009; Misra *et al.*, 2014a; Schwieterman *et al.*, 2015b). A planet with the appropriate temperature and pressure, in addition to the presence of H$_2$O absorption bands, is likely to be a habitable world (Des Marais *et al.*, 2002). Robinson (2017) provides a current review of habitability detection.

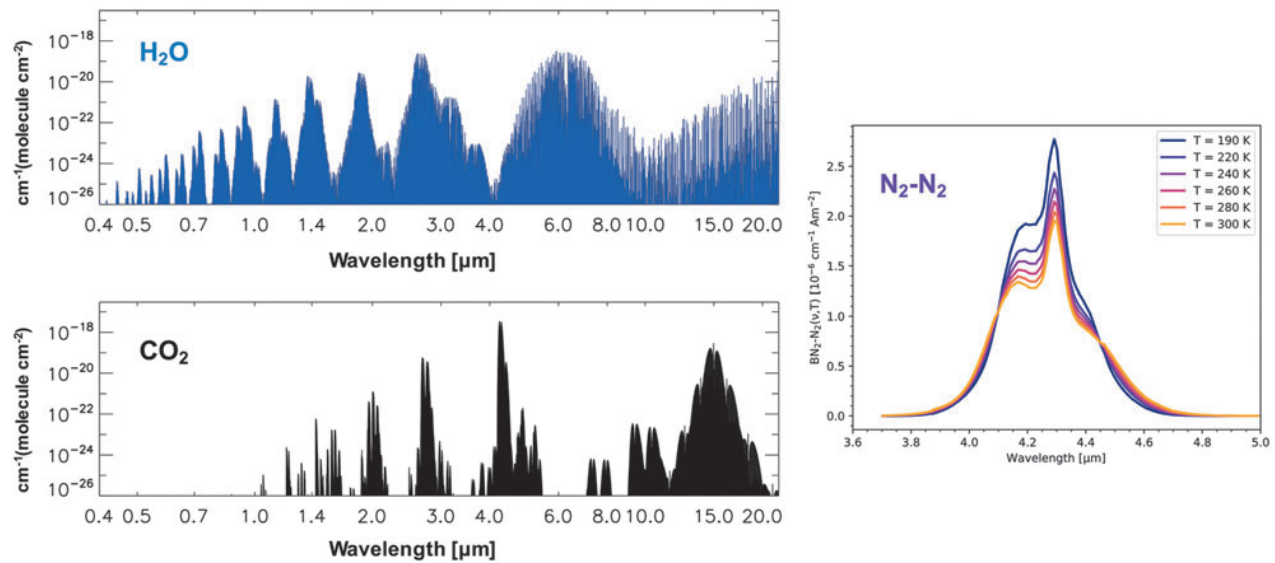

**FIG. 3.** Exoplanet habitability markers [H$_2$O + CO$_2$ + (N$_2$)$_2$]. Left: spectral line intensities for H$_2$O and CO$_2$ from the HITRAN 2012 line-by-line database (Rothman *et al.*, 2013). Right: temperature-dependent N$_2$-N$_2$ binary (collisional) absorption coefficients from a formulation by Lafferty *et al.* (1996), after a plot from Schwieterman *et al.* (2015b).

Conceptually, we can place potential exoplanet spectral habitability markers into the same broad categories as exoplanet biosignatures: gaseous, surface, and temporal. Water vapor and carbon dioxide gas would be examples of gaseous signatures (Fig. 3), ocean–continent heterogeneity and glint would be examples of surface signatures (although requiring a time component to the observation), while variable cloud cover and transient volcanic gases or aerosols are examples of temporal signatures of atmospheric properties that may be linked to habitability (*e.g.*, Kaltenegger *et al.*, 2010; Misra *et al.*, 2015). We leave focused and explicit exploration of habitability assessment and its relation to biosignature evaluation for the companion article in this issue by Catling *et al.* (2018; see especially their reference Tables 3 and 4). However, we continue to reference habitability markers here as they directly relate to biosignatures, such as in the case of spectral overlap of notable bands or feedback connections between habitability markers and biosignature gases in the atmosphere. We note that biosignatures must be examined in the context of effects and "background noise" due to putative habitability signatures (such as in the case of spectral overlap between, *e.g.*, H$_2$O and methane [CH$_4$]).

## 3. Overview of Terrestrial Exoplanet Modeling Studies

Because many potential biosignatures have been identified through models that variously treat the planetary atmospheric, biogeochemical, and physical systems, it is appropriate that we provide a short introduction to such modeling tools and studies. These strategies include data–model comparisons, photochemical models, spectral models, and studies of Earth's evolution. Such system-level approaches serve as frameworks and provide foundational concepts for discussions relating to exoplanet biosignatures.

### 3.1. Observations of Earth

As Earth currently offers our only example of an inhabited planet, observations of Earth have been analyzed for the detectability of biosignatures, and these data have then served to evaluate spectral models that simulate Earth radiance spectra for a variety of viewing geometries and cloud conditions. Two primary observing modes have provided data: (1) measurements of Earthshine reflected from the Moon (Arnold *et al.*, 2002; Woolf *et al.*, 2002; Montanes-Rodriguez *et al.*, 2005, 2006; Seager *et al.*, 2005; Hamdani *et al.*, 2006; Turnbull *et al.*, 2006; Arnold, 2008; Pallé *et al.*, 2009; Sterzik *et al.*, 2012) and (2) photometric and spectrophotometric observations of Earth by interplanetary spacecraft (Sagan *et al.*, 1993; Livengood *et al.*, 2011; Robinson *et al.*, 2011, 2014; Hurley *et al.*, 2014; Schwieterman *et al.*, 2015b). It is clear from these studies that habitability markers (H$_2$O, CO$_2$, N$_2$, and ocean glint, as described in the previous section), some biosignature gases (O$_2$, ozone [O$_3$]), and the vegetation red edge (VRE) surface biosignature can be detected in Earth's disk-averaged spectrum (these biosignatures are described in detail in sections 4 and 5 below). Heterogeneous features such as vegetation are more easily studied with significant spatial resolution (*e.g.*, Sagan *et al.*, 1993) or at opportune phases that maximize the viewable planetary surface through clear sky paths.

The first observations of potentially habitable exoplanets will likely be limited to disk-averaged photometry and spectra such that only those biosignatures with a global, planetary impact will be detectable. However, time-resolved photometry techniques have the potential to quantify heterogeneity of surface cover fractions on rapidly rotation planets (Ford *et al.*, 2001; Fujii *et al.*, 2010; Kawahara and Fujii, 2010; Fujii and Kawahara, 2012; Cowan and Strait, 2013). Biosignatures will have varying levels of detectability with different observing modes (*e.g.*, direct-imaging vs. transmission spectroscopy).

## 3.2. Spectral models

Radiative transfer models allow us to calculate the scattering and absorption of radiation through a medium such as an atmosphere, a body of water, or even a plant canopy. Such models are used to generate synthetic spectra of exoplanets, and are essential for estimating the remote detectability of biosignature gases or surface features. The type of radiative transfer model will vary depending on the planned observing mode to be simulated. Synthetic direct-imaging models simulate the reflected and emitted light from a planetary body. The reflected light includes incident stellar radiation that is scattered (or specularly reflected) to the observer by the planet's atmosphere or its surface. Emitted light is the thermal radiation from the planet. Transmission spectroscopy models simulate the spectrum of light that has passed through the atmosphere of a transiting exoplanet. Typically, high-resolution spectra require line-by-line approaches—for example, the Spectral Mapping Atmospheric Radiative Transfer (SMART) Model (Meadows and Crisp, 1996; Crisp, 1997) or the Generic Atmospheric Radiation Line-by-line Infrared Code (GARLIC) (Schreier *et al.*, 2014). In general, a line list database such as HITRAN (Rothman *et al.*, 2013), HITEMP (Rothman *et al.*, 2010), or ExoMol (Tennyson and Yurchenko, 2012) is used to query line parameters for gases included in the model and calculate absorption cross sections.

Separate modules of the radiative transfer model must calculate Rayleigh and aerosol scattering. Aerosol parameters (*e.g.*, particle size distributions, densities, and altitudes) must also be specified if haze or cloud cover is assumed. Surface spectral albedos constitute the lower boundary conditions in the spectral model and can be assumed to be Lambertian or the entire bidirectional reflectance distribution function can be specified. Spectral models are necessary for our understanding of exoplanet biosignatures because they must be used to determine whether a proposed biosignature gas, surface feature, or temporal modulation produces a sufficient impact to be detectable.

As an example of a spectral model of Earth validated through observations, Fig. 4 shows a simulated ultraviolet, visible, near-infrared, and mid-infrared (UV-VIS-NIR-MIR) spectrum of Earth from the well-validated Virtual Planetary Laboratory (VPL) 3D spectral Earth model (Robinson *et al.*, 2011). This model includes gaseous absorption, Rayleigh scattering, the modern Earth's actual continental and surface distribution, and realistic cloud cover. The model validation included the following: (1) data–model comparisons with visible spectrophotometric measurements and NIR spectroscopy by the EPOXI mission (Livengood *et al.*, 2011; Robinson *et al.*, 2011; Schwieterman *et al.*, 2015b), (2) MIR data–model comparisons with measurements from the Aqua Earth observing satellite, and (3) VIS-NIR spectra taken by the Lunar Crater Observation and Sensing Satellite (LCROSS) mission (Robinson *et al.*, 2014). Biosignature gas absorption features are present, including those of $O_3$, $O_2$, and $CH_4$. The VRE is included as well as water vapor absorption.

Data–model comparisons have the capacity to validate the detectability of biosignature features through forward modeling (*e.g.*, Des Marais *et al.*, 2002), providing these signatures exist on the modern Earth and their presence is in some way imprinted onto our planet's spectrum. Once validated, those same models can then be altered to simulate planetary spectra for different viewing geometries, cloud conditions, and alternative atmospheric compositions and surface features—in other words, for a wide array of planetary conditions. Through this approach, we can surmise the detectability of biosignatures, including biosignatures proposed for planets such as early Earth (Section 3.4), planets orbiting different stars, or biospheres under different environmental conditions. Sections 4, 5, and 6 describe in more detail studies of individual gaseous, surface, and temporal biosignatures, respectively, many of which use Earth model validations to confirm that spectral models can accurately represent the impact of a proposed biosignature on a planetary spectrum.

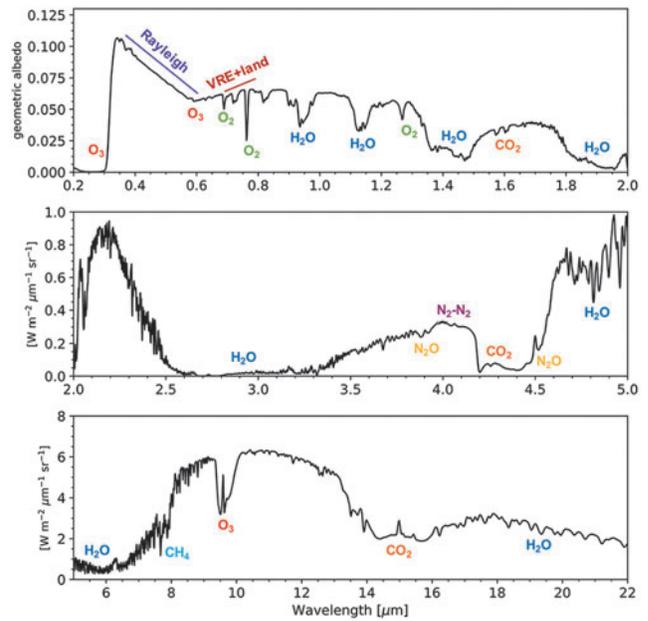

**FIG. 4.** A synthetic UVOIR Earth radiance spectrum at quadrature phase (half illumination). The top panel ($0.2\,\mu m < \lambda < 2.0\,\mu m$) is shown in terms of geometric albedo, while the bottom two panels ($2\,\mu m < \lambda < 5\,\mu m$; $5\,\mu m < \lambda < 22\,\mu m$) are presented as spectral radiance units ($W \cdot m^{-2} \cdot \mu m^{-1} \cdot sr^{-1}$). This spectrum was generated by the VPL 3D spectral Earth model (Robinson *et al.*, 2011; Schwieterman *et al.*, 2015b). Strong absorption features from $O_2$, $O_3$, $H_2O$, $CO_2$, $N_2O$, and $CH_4$ are labeled, in addition to Rayleigh scattering and the location of the VRE. 3D, three-dimensional; $CH_4$, methane; $N_2O$, nitrous oxide; UVOIR, ultraviolet-optical-infrared; VPL, Virtual Planetary Laboratory; VRE, vegetation red edge.

## 3.3. Photochemical studies of terrestrial atmospheres

Photochemical models simulate the interaction of a host star's radiation with a planetary atmosphere. These models use an abbreviated selection of chemical species and reactions to approximate the chemical composition of planetary atmospheres, which may then serve as input for spectral models. The list of species can include both gas-phase and aerosol-phase molecules. These models also track radiative transfer through the atmosphere, focusing on the UV to visible part of the spectrum, as photons in that wavelength range drive most photolysis reactions.

Generally, photochemical models calculate the rate of each reaction in the model (including but not limited to photolysis reactions) as well as physical mixing between model grid points and depositional mass fluxes. This combination allows the creation of a set of equations for the production and loss of each species in each layer of the model atmosphere. Together, these equations define a set of coupled differential equations that are passed to a numerical solver used to find a self-consistent solution for the atmospheric state based on the list of chemical species, their reactions, the stellar irradiation, and the assumed boundary conditions for the model grid.

The main boundary conditions for these models are as follows: (1) the mass fluxes into or out of the atmosphere (usually into the atmosphere at the surface–atmosphere interface, along with a limited flow of light species such as H out of the top of the atmosphere) and (2) the spectral energy flux into the top of the atmosphere, according to the star and the star–planet separation. These boundary conditions can fundamentally alter the composition of the atmosphere. Depending on the purpose and complexity of the photochemical model, it may ultimately calculate a steady-state atmospheric composition that is stable over geological timescales. Alternatively, some photochemical modeling efforts have focused on characterizing the atmospheric consequences of short-duration events, such as stellar flares, by using the same numerical tools.

There are several well-established photochemical models developed by different research groups over the last few decades. One of the Kasting group (e.g., Kasting, 1982, 1997; Pavlov and Kasting, 2002; Domagal-Goldman et al., 2011) and versions developed therefrom (e.g., Segura et al., 2003, 2007, 2010; Rauer et al., 2011; Arney et al., 2016), the Caltech/JPL model (e.g., Allen et al., 1981; Nair et al., 1994; Yung et al., 1988; Gao et al., 2015), and the Hu group model (Hu et al., 2012, 2013; Hu and Seager, 2014) all share the same general approach to simulating photochemistry. As mentioned earlier, these models include atmospheric chemical reaction lists for the major and minor species and represent a set of partial differential equations governing the concentrations of those species. The models use these equations to evolve gas concentrations toward steady state. Boundary conditions, as mentioned previously, include the impact of planetary processes (e.g., volcanism) on the atmosphere.

Photochemical models have been used in a variety of contexts. They have a long history of modern Earth applications, from modeling the $O_3$ hole and the evolution of greenhouse gas concentrations, to understanding the fate of trace pollutants. In planetary science, these models are often used to interpret data from spacecraft observations, or to simulate data returns from future missions. They also have been used to help understand the atmospheres of early Earth and early Mars by delineating atmospheric states that are consistent with geological and geochemical data.

For exoplanets, photochemical models have been used to simulate potential chemical compositions of a wide variety of worlds, to either interpret observed transit spectroscopy data or to simulate future spectral data, including spectral biosignatures. If spectral simulations are desired, the outputs of a photochemical model are used as inputs to a more detailed radiative transfer model that generates the spectrum. Photochemical models are especially useful for helping understand the contextual information required to interpret a biosignature. Examples of such studies include investigations of the potential atmospheric composition of the Archean Earth (Kasting, 2001; Kharecha et al., 2005; Kaltenegger et al., 2007), of planets orbiting M dwarf stars that have low UV flux (e.g., Segura et al., 2005), of possible by-products of biogenic gases that could serve as biosignatures (e.g., Domagal-Goldman et al., 2011), and of the possibility of false positives from abiotic generation of biosignature gases in alternative planetary environments (see Section 4).

Caution should be taken when interpreting the results of a photochemical model that simulates an atmosphere without the aid of observational constraints. These simulations represent a plausible and sustainable atmospheric state, not necessarily its current chemical composition. A planet could have multiple stable states, given a single set of boundary conditions. Conversely, it is possible for different sets of boundary conditions to reproduce the same stable state. Changing the boundary conditions in the model may result in different sets of possible states. These plausible photochemical model solutions are useful for assessing which of these states may contain detectable biosignatures, as well as to motivate research to constrain better the modeled processes. A prime example is in ongoing research to explain the suite of Earth's different geochemical states through time.

### 3.4. Earth through time

Life and environment have coevolved on Earth for billions of years. The most significant biologically mediated change was the oxygenation of Earth's atmosphere due to the evolution of oxygenic photosynthesis (OP); in turn, high levels of $O_2$ at Earth's surface eventually allowed the emergence and proliferation of complex, animal life. Evidence for atmospheric $O_2$ first appeared in the rock record $\sim 2.3$ billion years ago (Gyr; e.g., Luo et al., 2016) during a relatively short interval of time referred to as the "Great Oxidation Event" (GOE) (e.g., Holland, 2002) (Fig. 5). Another series of shifts in atmospheric $O_2$ occurred during the late Proterozoic and the early Phanerozoic ($\sim 750$ to 460 million years ago). Each geological eon comprises a suite of differences not only in the oxidation state of the atmosphere but also in the composition of the biosphere. Thus, each provides a potential template/analogue for the spectral character of a biogeochemical state of a rocky terrestrial planet in the HZ of its star.

Nuanced interpretation of potential observations of "Alternative Earth" analogues must also consider the uncertainty and possible lack of detectability of the life-forms that may be only just emerging or that abound but in ways that insufficiently impact the atmosphere. For example, the date of the earliest emergence of $O_2$-evolving photosynthetic cyanobacteria is highly uncertain. $O_2$ may have been produced by cyanobacteria in the late Archean, but at a rate that could not yet produce a strong atmospheric signal (Lyons et al., 2014); indeed, before the GOE, atmospheric "whiffs" of $O_2$, and localized $O_2$ oases in the shallow ocean are inferred from the trace element and isotope records (Planavsky et al., 2014a).

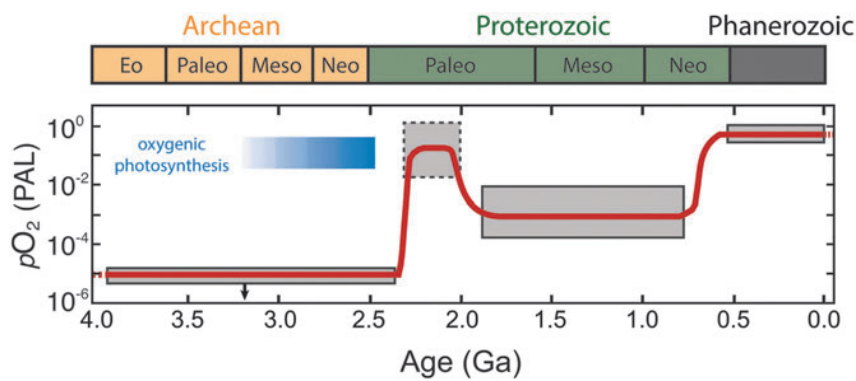

**FIG. 5.** Evolution of Earth's atmospheric $O_2$ content through time. Shaded boxes show approximate ranges based on the latest geochemical proxy records, while the curve shows one possible evolutionary trend over time. After Reinhard *et al.* (2017); see also Lyons *et al.* (2014).

Such traces of life would likely elude detection. (The mechanisms for $O_2$ buildup and further details surrounding the trajectory of the $O_2$ content of Earth's atmosphere are described in Section 4.2.1.)

Various modeling studies have attempted to model Earth's atmospheric composition and spectral signatures and their variation over time self-consistently by using coupled photochemical/radiative–convective models and prescribed surface fluxes of various gases or prescribed surface spectral albedos. Relevant examples include models exploring different geologic eons on Earth (Meadows, 2006; Kaltenegger *et al.*, 2007); an early Earth with photosynthetic microbial mats on land (Sanromá *et al.*, 2013), a purple Archean ocean due to photosynthetic purple bacteria (Sanromá *et al.*, 2014); an orange Archean Earth due to an organic-rich atmospheric haze (Arney *et al.*, 2016); and trajectories of $O_2$ fluxes over geologic time given estimated atmospheric concentrations (Gebauer *et al.*, 2017). Such approaches have also been applied to planets orbiting other stars: for example, biosignature gas concentrations and detectability under the UV environments of planets orbiting M stars (Segura *et al.*, 2005; Rugheimer *et al.*, 2015a, 2015b); and organic hazes on Earth-like planets around different stellar types (Arney *et al.*, 2017). The sections that follow describe these biosignature examples and others in detail, with many investigations in the context of photochemical models of alternative planetary scenarios.

## 4. Gaseous Biosignatures

Gaseous biosignatures can result from direct biological production or from environmental processing of biogenic products leading to secondary compounds. The example treated in detail in the companion article by Meadows *et al.* (2018) is $O_2$ produced from photosynthesis, and $O_3$ subsequently formed by photochemical reactions involving $O_2$ in the stratosphere. Not all biogenic gases are uniquely biological, and their identification as signs of life will depend strongly on their environmental context. Below we describe biogenic gases known to date, the contexts in which they may or may not be identified as biosignatures, their spectral absorbance features, and how they may be observed.

### 4.1. Gaseous biosignature overview

To be spectrally detectable, gases in the atmosphere must interact with photons through dissociation, electronic, or vibrorotational transitions. Because many gases absorb near the same wavelengths, it is essential to have the spectral range and/or resolution to discriminate between gases to uniquely identify their presence or absence in an exoplanet atmosphere. Figure 6 shows the line absorption intensities or absorption cross sections for the biosignature gases presented in this section for reference, drawing from the HITRAN 2012 (Rothman *et al.*, 2013) and PNNL (Sharpe *et al.*, 2004) spectral databases. These gases include $O_2$, $O_3$, nitrous oxide ($N_2O$), $CH_4$, methyl chloride ($CH_3Cl$), ethane ($C_2H_6$), $NH_3$, dimethyl sulfide (DMS), dimethyl disulfide (DMDS), and methanethiol ($CH_3SH$) (also see reference Tables 3 and 4 in Catling *et al.*, 2018, a companion article in this journal issue).

### 4.2. Earth-like atmospheres

An "Earth-like" atmosphere is defined here as one dominated by $N_2$, $CO_2$, and $H_2O$ ($O_2$ may or may not be a significant component). An "Earth-like" atmosphere is, by definition, associated with habitability and characterized by the presence of high-molecular-weight gases ($\mu_M \gg 2$) that include a condensable greenhouse gas ($H_2O$), a noncondensable greenhouse gas ($CO_2$), and a noncondensable background gas ($N_2$). This definition is traditionally used to define the circumstellar HZ with 1D radiative–convective climate models (Kopparapu *et al.*, 2013). Earth's atmospheric composition has evolved greatly through time (section 3.4), and so, it is important not to limit the definition of "Earth-like" to atmospheres identical to Earth's modern atmosphere, which represents just a small part of Earth history (*e.g.*, Lyons *et al.*, 2014). Furthermore, an Earth-like atmosphere is not the only type of "habitable" atmosphere conceivable for a rocky, terrestrial planet. Alternative possibilities, such as an $H_2$-dominated atmosphere, are described in Section 4.4.

Each subsection below describes a biosignature gas that has been considered for Earth-like atmospheres (high molecular weight, $N_2$-$CO_2$-$H_2O$ dominated). First, the major biological production and buildup mechanisms for the gas are described. Second, abiotic sources are presented and discussed if known. If the buildup of the gas has been studied as a function of the host star spectral type, this is also discussed. Each subsection concludes with a description of the major absorption bands of each gas, and whether they overlap with those from other gaseous biosignatures.

#### 4.2.1. Oxygen ($O_2$).
Molecular $O_2$ and its photochemical by-product $O_3$ have been the most highly referenced astronomical biosignature gases since surveys of nearby habitable planets have been contemplated (*e.g.*, Owen, 1980; Leger *et al.*, 1993; Sagan *et al.*, 1993; Des Marais *et al.*, 2002). This is largely because $O_2$ is a dominant gas in Earth's modern atmosphere ($pO_2 = 0.21$), produces potentially detectable spectral signatures, and is effectively entirely sourced from photosynthesis on Earth. Oxygenic photosynthesis (OP) uses light energy to (indirectly) split $H_2O$, which serves as an electron donor to produce organic matter from $CO_2$, generating $O_2$ as a waste product (Leslie, 2009). The net reaction is often written as follows:

$$CO_2(g) + 2H_2O^w + h\nu \rightarrow (CH_2O)_{org} + H_2O + O_2{}^w(g)$$

where $(CH_2O)_{org}$ represents organic matter and $h\nu$ is the energy of the photon(s) (where h is Planck's constant and $\nu$ is the frequency of the photon). Although the net equation may cancel an $H_2O$ from both sides, this representation explicitly shows that the $O_2$ atoms for the evolved $O_2$ (denoted with superscript $w$) come from the water molecules and not the carbon dioxide. OP makes use of some of the most widely available molecules in the ocean–atmosphere system ($H_2O$ and $CO_2$) and harnesses abundant photons from the Sun. It is regarded as perhaps the most potentially productive metabolism on any planet orbiting a star due to the wide availability of its basic substrate and energy source (Kiang *et al.*, 2007a, 2007b). The range of organisms that use OP on our own planet includes plants, algae, and cyanobacteria. It is important to note that oxygenation of an atmosphere is a more complex process than simple production of $O_2$ by photosynthetic organisms. The net photosynthetic reaction given above is, in a general sense, reversible, depleting $O_2$ with the decay of organic matter via aerobic respiration ($CH_2O + O_2 \rightarrow CO_2 + H_2O$). Photosynthesis produces no net $O_2$ unless some of the organic matter is preserved and ultimately sequestered from the atmosphere. This process is primarily facilitated by burial of organic matter in marine sediments or soils (Berner and Canfield, 1989; Bergman, 2004; Catling, 2014), and is also greatly augmented by burial of sulfide generated by anaerobic sulfate reducers that oxidize organic matter (Berner and Raiswell, 1983). The accumulation of $O_2$ in the atmosphere further requires that the rate of these burial processes is greater than the rate of $O_2$ losses, such as by reactions with reduced volcanic gases (Catling, 2014).

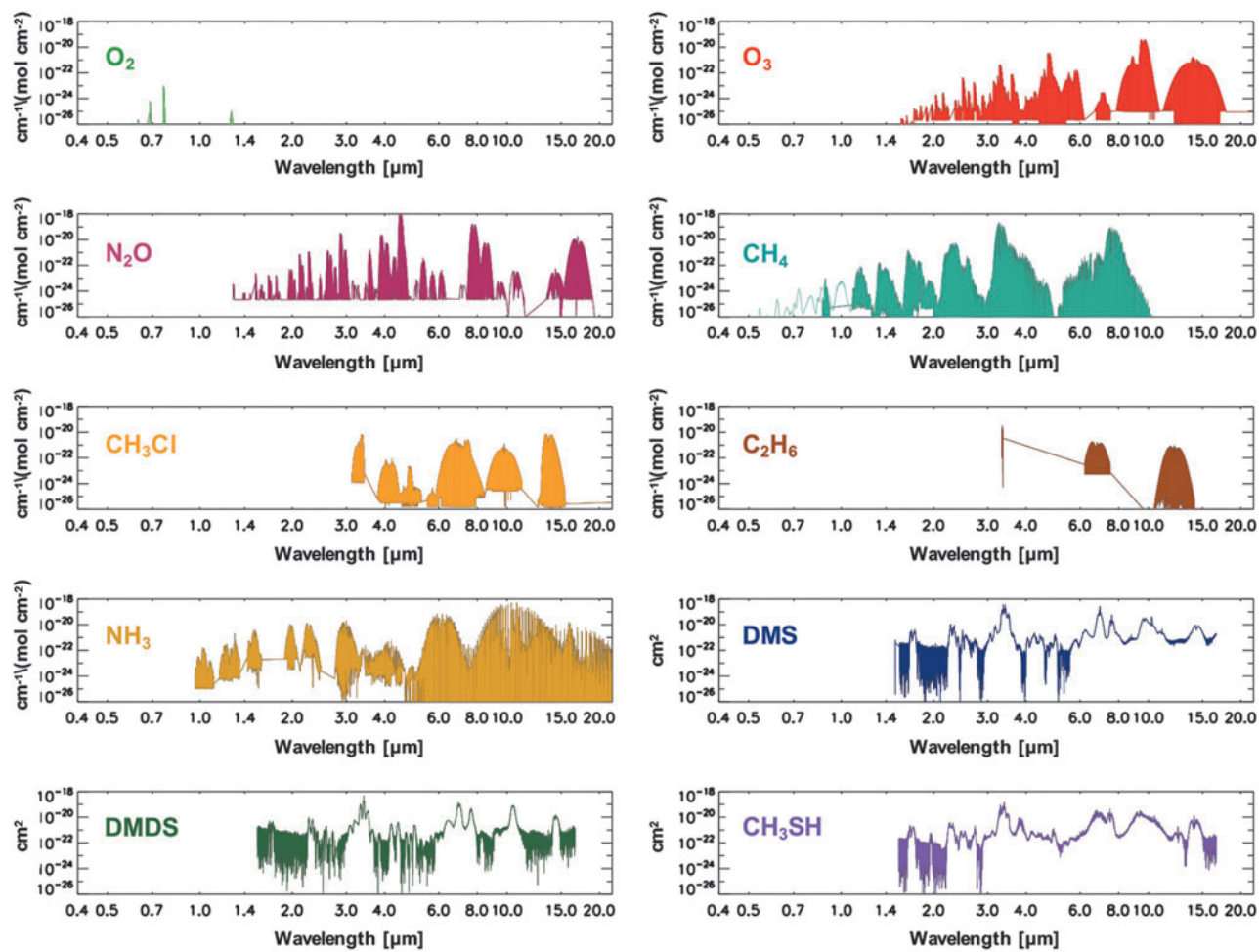

**FIG. 6.** Biosignature gas absorption features. Line intensities (cm⁻¹/[molecule cm⁻²]) for $O_2$, $O_3$, $N_2O$, $CH_4$, $CH_3Cl$, $C_2H_6$, and $NH_3$ are sourced from HITRAN 2012 (Rothman *et al.*, 2013), while cross sections (cm²) for DMS, DMDS, and $CH_3SH$ are sourced from the PNNL spectral database (Sharpe *et al.*, 2004). $C_2H_6$, ethane; $CH_3Cl$, methyl chloride; $CH_3SH$, methanethiol; DMDS, dimethyl disulfide; DMS, dimethyl sulfide.

The history of Earth's $O_2$ levels has many nuances, but there is a broad consensus on the major phases (*e.g.*, Lyons *et al.*, 2014). In the Archean eon (4.0–2.5 Ga), $O_2$ levels were very low ($pO_2 < 10^{-7}$), while $CH_4$ levels were believed to be elevated (100–1000 ppm). The GOE occurred ca. 2.4 Ga, near the beginning of the Proterozoic eon, and marked a significant change in the chemistry of the atmosphere, increasing $pO_2$ by several orders of magnitude and decreasing the prevalence of reduced gases such as $CH_4$. After the GOE $pO_2$ rose to as high as 1–10% of modern levels (Kump, 2008), although recent evidence suggests $pO_2$ remained low relative to modern levels for most of the Proterozoic eon ($pO_2 < 0.1\%$) (Planavsky *et al.*, 2014b), it was only after a second series of $O_2$ shifts during the late Proterozoic ($\sim 800$ million years ago, Ma) and the Paleozoic ($\sim 420$ Ma) when $pO_2$ approached modern levels. The late Proterozoic shift occurred roughly contemporaneously with the rise and diversification of complex animal life (Reinhard *et al.*, 2016). Importantly, the initial rise of $O_2$ levels on Earth was delayed until well after the evolution of OP, which had likely occurred in modern levels for most of the Proterozoic eon
if not earlier. In any case, understanding the protracted rise of $O_2$ in Earth's atmosphere is an active area of investigation, with critical implications for biosignature evolution on extrasolar planets.

Molecular $O_2$ has a few strong bands in the VIS/NIR region, including the $O_2$-A band (0.76 μm), the $O_2$-B band (0.69 μm), and the $O_2$-g band (0.63 μm). In addition, $O_2$ collisionally induced absorption ($O_2$-$O_2$) occurs at 1.06 μm, and the 1.27 μm $O_2$ band includes contributions both from monomer $O_2$ ($a^1\Delta_g$ band) and dimer $O_2$-$O_2$ collisional absorption. At very high $O_2$ concentrations, $O_2$-$O_2$ CIA (also referred to as $O_4$) absorption occurs at 0.445, 0.475, 0.53, 0.57, and 0.63 μm (Hermans *et al.*, 1999; Richard *et al.*, 2012; Schwieterman *et al.*, 2016). In the MIR, $O_2$ has an absorption band at 6.4 μm, but this band is weak and overlaps with much stronger $H_2O$ absorption, so is unlikely to be observable at low resolution for habitable planets. In the UV, $O_2$ has strong absorption from photodissociation at wavelengths shorter than 0.2 μm, although this is also true for several other gases, such as $CO_2$. Of these, the $O_2$-A band (0.76 μm) is by far the most preferable target band for direct imaging (reflected light observations) due to its relative strength and the lack of overlap with features from other common gases.

On Earth, the production of abiotic $O_2$ from the photolysis of other O-bearing molecules occurs at a very slow rate. This $O_2$ would not build up to appreciable levels due to the distribution of UV energy from the Sun (which controls the rate of $O_2$ production from O-bearing species such as $CO_2$ and its destruction rate) and geochemical sinks for $O_2$ (*e.g.*, Domagal-Goldman *et al.*, 2014; Harman *et al.*, 2015). However, several scenarios have been described that could allow for the buildup of abiotic $O_2$ for planets orbiting other types of stars. Potential "false positives" for abiotic $O_2$ are reviewed briefly in Section 4.3 and more extensively in Meadows (2017) and Meadows *et al.* (2018).

### 4.2.2. Ozone ($O_3$).

The $O_3$ in Earth's stratosphere is the result of photochemical reactions that split $O_2$. The detection of significant $O_3$ in a planetary atmosphere has been proposed as a proxy for photosynthetically generated $O_2$

(Léger *et al.*, 1993, 2011; Des Marais *et al.*, 2002), with the advantage that $O_3$ absorbs strongly in complementary wavelength bands to $O_2$ (*e.g.*, in the UV and MIR). The formation and destruction cycle of $O_3$ is described by the Chapman scheme (Chapman; 1930):

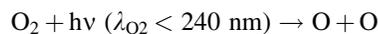

$$O_2 + h\nu\ (\lambda_{O2} < 240\ nm) \rightarrow O + O$$

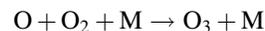

$$O + O_2 + M \rightarrow O_3 + M$$

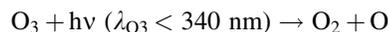

$$O_3 + h\nu\ (\lambda_{O3} < 340\ nm) \rightarrow O_2 + O$$

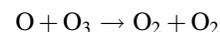

$$O + O_3 \rightarrow O_2 + O_2$$

where $\lambda$ is the minimum wavelength for photodissociation of the given molecule and M is any molecule that can carry away excess vibrational energy (*e.g.*, $N_2$). The $O_3$ layer on Earth reaches peak concentrations of up to 10 ppm in the stratosphere between 15 and 30 km in altitude, but both the value and altitude of the peak $O_3$ concentration vary spatially. The incident UV photon flux and spectrum impact the rate of $O_3$ production and destruction, thus affecting the predicted $O_3$ concentration and profile for planets orbiting different stars even if the planetary $O_2$ abundance is the same (Segura *et al.*, 2003; Rugheimer *et al.*, 2013; Grenfell *et al.*, 2014). Indeed, planets with the same $O_2$ abundances, orbiting the same star, but at different distances, will have slightly different $O_3$ profiles mainly due to differences in UV and temperature structure (Grenfell *et al.*, 2007). Furthermore, particle fluxes from flares around active stars have the capacity to strongly attenuate the predicted $O_3$ column, depending on the strength and frequency of the flare events (Segura *et al.*, 2010; Tabataba-Vakili *et al.*, 2016). Like $O_2$, $O_3$ may be produced through abiotic photochemical mechanisms, with current literature studies indicating that abiotic production is favored most around M dwarf and F dwarf stars (Domagal-Goldman *et al.*, 2014; Harman *et al.*, 2015). This relationship is further discussed in Section 4.4.

$O_3$ possesses absorption features in the UV-VIS-NIR-MIR regions of the spectrum. In the UV, the Hartley–Huggins bands are centered at 0.25 μm and extend from 0.35 to 0.15 μm. These bands are saturated in Earth's spectrum, but caution is warranted since other molecules such as sulfur dioxide ($SO_2$) also absorb in this wavelength region (Robinson *et al.*, 2014). In the visible, the Chappuis bands extend from 0.5 to 0.7 μm and contribute to the "U" shape of Earth's overall UV-VIS-NIR spectrum, a feature that distinguishes Earth's color from those of other planets at even very low ($\Delta\lambda \sim 0.1$ μm) spectral resolution (Krissansen-Totton *et al.*, 2016b). $O_3$ has several weak bands in the NIR at 2.05, 2.15, 2.5, 3.3, 3.6, 4.6, and 4.8 μm. Those at the longer wavelengths are the strongest, although many of these bands overlap with absorption features from $H_2O$ and $CO_2$. In reflected light, the UV band is the strongest feature from $O_3$. $O_3$ also imprints strong features on the emitted, thermal infrared portion of Earth's spectrum. The strongest and most well studied of these is the 9.65 μm band, which occurs in the middle of Earth's thermal infrared spectral window (Des Marais *et al.*, 2002). The 9.65 μm band would be a prime target for an infrared-capable telescope, such as the previously envisioned Terrestrial Planet Finder–Infrared mission (TPF-I) (Beichman *et al.*, 2006; Lawson *et al.*, 2006; Traub et al., 2007) or its ESA equivalent Darwin (Cockell et al., 2009b). Caution should be given to the overlap from the "doubly hot" band of $CO_2$ at 9.4 µm, which would also produce absorption at 10.5 µm (Segura et al., 2007, see their Fig. 5b). In addition, other gases, including $CH_3Cl$, DMS, DMDS, and $CH_3SH$, have overlapping absorption features near the 9.65 µm band (Pilcher, 2003, see sections 4.2.5 and 4.2.6 below). Therefore, it is essential to obtain spectral information at other wavelengths to confidently detect $O_3$. Finally, there is a weak $O_3$ band at 14.08 µm, which is completely swamped by the 15 µm $CO_2$ band. To summarize, the best prospects for detecting $O_3$ are the Hartley–Huggins bands centered at 0.25 µm in the UV, the subtler Chappuis band extending from 0.5 to 0.7 µm in the visible, and the 9.65 µm band in the MIR.

### 4.2.3. Methane ($CH_4$).

Methanogenesis is an ancient form of anaerobic microbial metabolism that produces $CH_4$ as a waste product, most commonly by either respiring $CO_2$ as a terminal electron acceptor or disproportionating acetate to $CH_4$ and $CO_2$. These reactions can be written as follows:

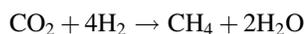

$$CO_2 + 4H_2 \rightarrow CH_4 + 2H_2O$$

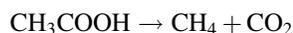

$$CH_3COOH \rightarrow CH_4 + CO_2$$

where $H_2$ is hydrogen gas and $CH_3COOH$ is acetic acid—a decay product from fermentation of organic matter. On Earth, the single-celled organisms responsible for methanogenesis, called "methanogens," are restricted to the domain Archaea. Methanogenesis is the dominant source of nonanthropogenic $CH_4$ in Earth's modern atmosphere, and $CH_4$ has consequently been suggested as a potential biosignature on Earth (e.g., Sagan et al., 1993) and on Mars (e.g., Krasnopolsky et al., 2004). However, there are many potential abiotic $CH_4$ sources, almost all of which involve water–rock reactions. See Etiope and Sherwood-Lollar (2013) for a review of abiotic $CH_4$ sources on Earth.

Primitive planet-building material from the outer Solar system, beyond the ice line, is replete with $CH_4$, since it is the most thermodynamically stable form of carbon in reducing (i.e., H-abundant) conditions. Therefore, planetary bodies constructed from this material may be expected to contain an abundance of abiotic $CH_4$. Such is the case in the atmosphere of Saturn's icy moon Titan, whose atmosphere contains 5% $CH_4$ by volume. $CH_4$ is likewise the most thermodynamically stable form of carbon in highly reducing, $H_2$-dominated atmospheres. Therefore, $CH_4$ is often viewed as a companion biosignature that would be most compelling if observed together with $O_2/O_3$ or other strongly oxidizing gases. $CH_4$ may also serve as a biosignature or habitability marker with the presence of $CO_2$, since the presence of $CO_2$ implies the atmosphere's redox state is more oxidizing and thus not conducive to producing $CH_4$ as the most stable form of carbon (Titan's atmosphere has very little $CO_2$). In an atmosphere with a significant amount of $CO_2$, the $CH_4$ would have had to originate from biology or from abiotic water–rock reactions, an indirect evidence of liquid water in the planetary environment.

The dominant sinks for $CH_4$ in the modern Earth's atmosphere involve oxidation of $CH_4$ by radical species, such as hydroxyl (OH), O($^1$D), or Cl, for example:

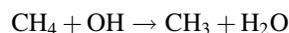

$$CH_4 + OH \rightarrow CH_3 + H_2O$$

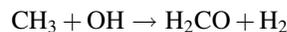

$$CH_3 + OH \rightarrow H_2CO + H_2$$

Formaldehyde ($H_2CO$) formed through this reaction can be further oxidized to $CO_2$ and $H_2O$, or incorporated into rain and transported to the ocean. In more reducing atmospheres, $CH_4$ photodissociation can drive the formation of longer chain hydrocarbons, ultimately leading to organic haze particles, as observed in Titan's atmosphere. Under anoxic conditions, $CH_4$ tends to be long lived in the atmosphere, but the advent of OP on Earth eventually led to the dramatic reduction of atmospheric $CH_4$ content (Pavlov and Kasting, 2002).

$CH_4$ absorbs throughout the VIS-NIR-MIR with its strongest bands at 1.65, 2.4, 3.3, and 7–8 µm. There are also weaker bands at (in order of increasing strength) 0.6, 0.7, 0.8, 0.9, 1.0, 1.1, and 1.4 µm. However, $CH_4$ bands in the visible and NIR are relatively weak at the abundances of modern Earth. The strongest band, centered between ∼7 and 8 µm, absorbs at the long-wavelength wing of the ∼6 µm $H_2O$ band and overlaps with $N_2O$, which also absorbs strongly between 7 and 9 µm. At each of $CH_4$'s strong bands, it overlaps with $H_2O$ absorption, which makes uniquely detecting $CH_4$ problematic at low spectral resolution.

### 4.2.4. Nitrous oxide ($N_2O$).

$N_2O$ is produced by Earth's biosphere via incomplete denitrification of nitrate ($NO_3^-$) to $N_2$ gas. A simplified scheme for denitrification can be written as follows:

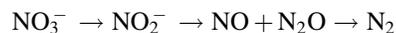

$$NO_3^- \rightarrow NO_2^- \rightarrow NO + N_2O \rightarrow N_2$$

$N_2O$ has been proposed as a strong biosignature, in part, because its abiotic sources are small on modern Earth and because it has potentially detectable spectral features (Sagan et al., 1993; Segura et al., 2005; Rauer et al., 2011; Rugheimer et al., 2013, 2015a). The preindustrial concentration of $N_2O$ in Earth's atmosphere was ∼270 ppb (Myhre et al., 2013). It has been proposed that euxinic oceans (replete with hydrogen sulfide [$H_2S$]) during portions of the Proterozoic epoch would have stifled the bioavailability of copper that facilitates the last step in the denitrification process (i.e., the reduction of $N_2O$ to $N_2$), allowing biogenic $N_2O$ to build up in the atmosphere with possible climatic implications (Buick, 2007; Roberson et al., 2011). For a biogeochemically analogous world, $N_2O$ may exist at higher concentrations than seen on modern Earth (e.g., Meadows, 2006; Kaltenegger et al., 2007). Photochemical modeling of terrestrial atmospheres around M dwarf stars has shown that $N_2O$ would build up to higher concentrations than on an Earth–Sun analogue given the same source fluxes (Segura et al., 2005; Rauer et al., 2011; Rugheimer et al., 2015a). This is a mechanism like that described for $CH_4$ above (Section 4.2.3) and is due, in large part, to a paucity of near-UV photons from cool M stars owing to their lower effective temperatures.

A small abiotic source of $N_2O$ on Earth is known from "chemodenitrification" of dissolved nitrates in hypersaline ponds in Antarctica (Samarkin et al., 2010), although the synthesis of $NO_3^-$ requires photosynthetically generated $O_2$. In this scenario, therefore, abiotic $N_2O$ production is ultimately an expression of biological activity on Earth. A small amount of $N_2O$ is also produced by lightning (Levine *et al.*, 1979), although the estimated contribution of total atmospheric $N_2O$ from lightning on Earth is 0.002% (Schumann and Huntrieser, 2007). Around young or more magnetically active stars, $N_2O$ may build up abiotically due to enhanced production of NO and NH from extreme ultraviolet (EUV-XUV) and particle flux-induced photodissociation and ionization, driving the reaction NO + NH $\rightarrow N_2O + H$ (Airapetian *et al.*, 2016). However, abiotic processes that generate $N_2O$ create associated nitrogen oxide ($NO_x$) products in far greater abundance than $N_2O$ (Schumann and Huntrieser, 2007), some of which may be spectrally observable and thus provide a signature of this process. In contrast, cosmic ray events are predicted to destroy $N_2O$ and favor production of nitric acid ($HNO_3$) (Tabataba-Vakili *et al.*, 2016).

Ultimately, the confidence with which $N_2O$ can be considered a robust biosignature must be evaluated in the context of the stellar environment as well as through observation of other photolytic products that would indicate abiotic $N_2$ oxidation. Fortunately, current studies suggest that abiotic sources of $N_2O$ are small except in cases where its production would be contextually predictable or inferable from planetary observations at many wavelengths, although disentangling $N_2O$ spectral features from overlapping gases may be difficult at low to moderate spectral resolving powers. $N_2O$ has significant bands centered at 3.7, 4.5, 7.8, 8.6, and 17 μm, with several weak bands between 1.3 and 4.2 μm and between 9.5 and 10.7 μm. However, most of these bands are weak at Earth-like abundances and/or overlap with other potentially abundance gases such as $H_2O$, $CO_2$, or $CH_4$, which may make detecting $N_2O$ challenging (Fig. 4). Observations at very high spectral resolution powers, at the level required to identify individual lines, may allow unique detection of $N_2O$ from overlapping gas absorption features.

### 4.2.5. Sulfur gases (DMS, DMDS, $CH_3SH$) and relation to detectable $C_2H_6$.

Biology produces several sulfur-bearing gases as direct or indirect products of metabolism. The direct products of metabolism tend to be simple sulfur gases such as $H_2S$, carbon disulfide ($CS_2$), carbonyl sulfide (OCS), and $SO_2$, although these are also produced in abundance by abiotic volcanic and hydrothermal processes and thus are not strong biosignature gas candidates [*e.g.*, see Arney *et al.* (2014), for an analysis of these gases in the Venusian atmosphere]. More complex sulfur gases such as $CH_3SCH_3$ or DMS, $CH_3S_2CH_3$ or DMDS, and $CH_3SH$ (also known as methyl mercaptan) are produced as indirect products of metabolism but have few if any known abiotic sources on modern Earth.

The organosulfur gases ($CH_3SH$, DMS, DMDS) are produced by bacteria and higher order life-forms in a variety of environments, including wetlands, inland soils, coastal ecosystems, and oceanic environments (Rasmussen, 1974; Aneja and Cooper, 1989). There are two principal routes to the production of DMS. The first is the biological degradation of the compound dimethylsulfoniopropionate (DMSP), which is found primarily in eukaryotic organisms such as certain types of marine algae (Stefels *et al.*, 2007). This pathway is believed to be the dominant source of DMS, which is the largest source of organosulfur gas in the modern atmosphere (Stefels *et al.*, 2007). Second, DMS

(and DMDS) can ultimately result from the production of $CH_3SH$, itself a decomposition product of the essential amino acid methionine. Microbial mats containing cyanobacteria and anoxygenic phototrophs produce measurable amounts of $CH_3SH$, DMS, and DMDS (Visscher *et al.*, 1991, 2003), likely from the reaction of short-chain organic compounds produced by the phototrophs reacting with sulfide produced by sulfate-reducing bacteria.

It has been hypothesized that the more reducing environment of early Earth would have been conducive to the production of greater volumes of sulfur gases by the anoxic biosphere (Pilcher, 2003; Domagal-Goldman *et al.*, 2011). The potential for photochemical buildup and the detectability of sulfur gases on early Earth exoplanet analogues were investigated by Domagal-Goldman *et al.* (2011), who considered biospheres that produced between 1 and 30 times the estimated modern-day fluxes for these gases during the Archean eon. These authors found that DMS, DMDS, and $CH_3SH$ were rapidly destroyed by photolysis reactions in the atmosphere, leading to near-zero mixing ratios at all but the lowest levels of the atmosphere (*e.g.*, see Domagal-Goldman *et al.*, 2011; their Fig. 2). Moreover, that study found that, even for biospheres with very high sulfur fluxes, their low abundance in the atmosphere, a consequence of efficient photochemical destruction, would render DMS, DMDS, and $CH_3SH$ spectrally undetectable except in the narrow case of an M star with suppressed UV activity (Domagal-Goldman *et al.*, 2011). However, the study also found that the cleaving of methyl ($CH_3$) radicals from DMS and DMDS by UV radiation catalyzed the photochemical buildup of $C_2H_6$ far beyond the level expected for the abundance of $CH_4$ in the atmosphere, which is otherwise the primary precursor to $C_2H_6$. Consequently, it was proposed that this anomalously high $C_2H_6$ signature would be suggestive of a sulfur biosphere (Domagal-Goldman *et al.*, 2011), although this may only be the case for high flux rates of organosulfur gases in combination with stellar hosts with a favorable UV spectrum for $C_2H_6$ production.

However, the detection of $C_2H_6$ alone would be an ambiguous signature, since photochemical processing of other carbon-bearing species such as $CH_4$ can generate it. The link to an organosulfur biosphere would necessitate constraints on the $C_2H_6$ to $CH_4$ abundance ratio to determine whether there is an overabundance of $C_2H_6$ relative to that which would be expected only from photochemical equilibrium with the retrieved $CH_4$ abundance. This would reveal the likelihood of other sources of $CH_3$ such as DMS that would act to increase the amount of $C_2H_6$. Necessarily, this comparison would also require forward modeling of the atmospheric photochemistry given the UV spectrum of the host star, the retrieved $CH_4$ abundance, other measured or likely atmospheric constituents, and additional planetary parameters such as the atmospheric temperature structure, which may or may not be available.

Although Domagal-Goldman *et al.* (2011) evaluated only synthetic direct-imaging spectra in their investigation of DMS and DMDS spectral detectability, their results also apply to transmission spectroscopy. Although transmission spectroscopy can enhance the signature of gases with low abundances through path length effects (*e.g.*, Fortney, 2005), this advantage is relevant only for gases with a presence in the portions of the atmosphere probed by transmission spectroscopy. Due to the combined effects of refraction, clouds, and aerosols, the lowest levels of an Earth-like atmosphere are not accessible (García Muñoz *et al.*, 2012; Bétrémieux and Kaltenegger, 2014; Misra *et al.*, 2014a, 2014b). Domagal-Goldman *et al.* (2011) found DMS and DMDS drop to near-zero abundance at all but the lowest levels of the atmosphere. Combined with results from direct imaging, these relationships suggest DMS and DMDS are examples of gases that, while exhibiting measurable spectral signatures in a laboratory setting, may never reach the required abundances to be directly detectable over interstellar distances for plausible biospheres. More encouragingly, their presence may be indirectly inferred by the detection of their photochemical by-products, in this case $C_2H_6$. This approach is analogous to the detection of $O_3$ to infer the presence of its primary precursor, $O_2$, in the atmosphere, with appropriate caveats considering other photochemical sources of $C_2H_6$ stated previously.

The strongest features of DMS are in the MIR at 6–7, ∼10, and ∼15 µm. DMDS absorbs strongest spectrally in the MIR at 7, 8–9, and 17 µm. $CH_3SH$ has its strongest features at 6–7, 8–12, and 14–15 µm. Notably, these gases all have absorption features that overlap with the 9.65-µm $O_3$ band (Pilcher, 2003), which could be problematic at low spectral resolution. $C_2H_6$ has strong spectral signatures at 6–7 and 11–13 µm and a weaker band at 3.3 µm.

### 4.2.6. Methyl chloride ($CH_3Cl$).

$CH_3Cl$, or chloromethane, is a biogenic gas whose major sources on Earth are both natural and anthropogenic: algae in the oceans (Singh *et al.*, 1983; Khalil and Rasmussen, 1999), tropical/subtropical plants (Yokouchi *et al.*, 2002, 2007; Saito and Yokouchi, 2006), aquatic plants in salt marshes (Rhew *et al.*, 2003), terrestrial plants (Saini *et al.*, 1995; Rhew *et al.*, 2014), fungi (Harper, 1985; Watling and Harper, 1998), decay of organic matter (Keppler *et al.*, 2000; Hamilton *et al.*, 2003), biomass burning (Lobert *et al.*, 1999), and industrial processes involving organic matter (Kohn *et al.*, 2014; Thornton *et al.*, 2016). Volcanoes may be an abiotic source (Schwandner *et al.*, 2004; Frische *et al.*, 2006). The relative contributions of these biological and abiotic sources remain unknown for the modern and ancient Earth (Keene *et al.*, 1999). The biological production mechanisms for $CH_3Cl$ are also poorly characterized (Rhew *et al.*, 2014), although a $CH_3$ chloride transferase enzyme has been identified (Ni and Hager, 1998), and methylation of plant pectin (during degradation) is a general pathway across taxa (Hamilton *et al.*, 2003). It appears there is not a unique pathway to production, but biosynthesis in numerous organisms, decay or combustion of organic matter, and volcanic gas-phase reactions can all produce $CH_3Cl$.

Spectral absorbance features occur at 3.3, 7, 9.7, and 13.7 µm (Rothman *et al.*, 2013) (note overlap with the 9.65 µm $O_3$ band). The dominant pathway for removal of $CH_3Cl$ is reaction with OH radical, with an estimated atmospheric lifetime of 1.3 years on Earth (WMO, 2003). In stellar environments with extremely low NUV flux suppressing OH formation, such as would take place in the quietest thermal lower limit of M dwarf activity with no chromospheric excess UV flux, there is a potential to build up $CH_3Cl$ to detectable levels (Segura *et al.*, 2005), although the feature overlaps with other expected features such as $H_2O$, $CO_2$, $O_3$, and $CH_4$.

$CH_3Cl$ would be best observed at 13.7 µm in the wings of the $CO_2$ feature (Rugheimer *et al.*, 2015a).

### 4.2.7. Haze as a biosignature.

Geochemical evidence suggests the existence of an intermittent organic haze during the late Archean geological eon (Zerkle *et al.*, 2012; Izon *et al.*, 2017). This haze would have dramatically impacted Earth's climate, photochemistry, and spectral observables. The putative Archean organic haze is like the organic haze in Titan's atmosphere in that it likely forms from $CH_4$ photochemistry. At the $CO_2$ levels suggested for the Archean Earth (atmospheric fractions of roughly $10^{-3}–10^{-2}$) (Driese *et al.*, 2011), a ratio of $CH_4/CO_2$ of about 0.2 is required to initiate the formation of a thick organic haze (*e.g.*, Trainer *et al.*, 2006). $CH_4$ on Earth can be produced by both abiotic and geological processes. On the modern Earth, biological processes produce the bulk of the atmosphere's $CH_4$, as was likely during the Archean eon (Kharecha *et al.*, 2005). The dominant abiotic $CH_4$ source on modern Earth—and likely the dominant abiotic source during the Archean—is serpentinization, the hydration of ultramafic minerals such as olivine and pyroxene (Kelley, 2005; Etiope and Sherwood-Lollar, 2013; Guzmán-Marmolejo *et al.*, 2013), although the ultimate source of $CH_4$ in serpentinizing systems is not entirely clear (McDermott *et al.*, 2015; McCollom, 2016).

Coupled photochemical-climate modeling has shown that producing a thick organic haze in the atmosphere of an exoplanet with Archean Earth-like $CO_2$ levels requires surface $CH_4$ flux rates consistent with measured modern biological $CH_4$ production [roughly $10^{11}$ molecules/(cm²·s)] and theoretical Archean biological $CH_4$ production rates [∼0.3–2.5 × $10^{11}$ molecules/(cm²·s), Kharecha *et al.*, 2005; also see Arney *et al.*, 2016, 2017, 2018]. Like $CH_4$ itself, organic haze would not definitively imply the existence of life, but organic haze produces strong, broadband absorption features at UV-blue wavelengths (the reason why Titan is orange), potentially more detectable than $CH_4$ itself. Because haze dramatically alters the broadband shape of a planet's reflected light spectrum (Fig. 7), it may provide a relatively simple means of flagging interesting targets for follow-up observations to search for other signs of habitability and life. Organic haze formation is incompatible with $O_2$-rich atmospheres and so would exist exclusively on exoplanets with reducing atmospheres, providing a useful means for identifying potentially inhabited worlds with more reducing atmospheres than modern-day Earth.

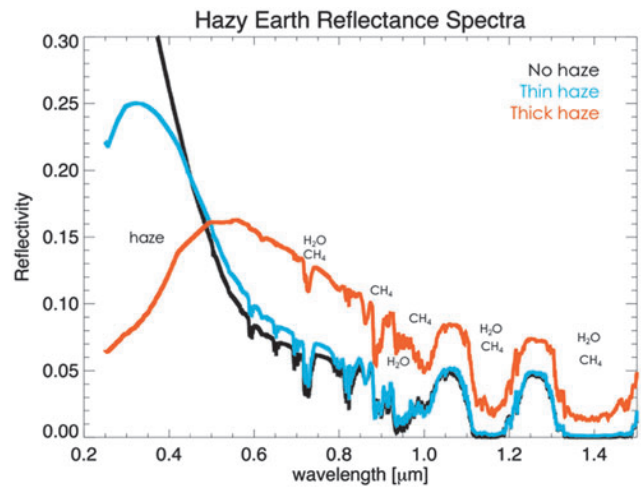

**FIG. 7.** Spectra of Archean Earth with three different haze thicknesses for atmospheres with 2% $CO_2$ (Arney *et al.*, 2016). The haze absorption feature at UV-blue wavelengths is strong and potentially detectable at spectral resolving powers as low as 10. UV, ultraviolet.

In the same way that organic haze could serve as an indicator of the $CH_4/CO_2$ ratio, and therefore, a gauge of the $CH_4$ flux, it has been proposed that sulfur aerosols ($S_8$ and $H_2SO_4$) could serve as a proxy of the $H_2S/SO_2$ ratio and a gauge of the $H_2S$ flux (Hu *et al.*, 2013). At high $H_2S/SO_2$ flux ratios, and a neutral to reducing atmosphere, $S_8$ aerosols would be formed. At low $H_2S/SO_2$ flux ratios, $H_2SO_4$ is formed preferentially over $S_8$. Oxidizing conditions that include even trace amounts of $O_2$ ($<10^{-5}$ present atmospheric level [PAL] $pO_2$) result in $H_2SO_4$ formation dominating over $S_8$ (Pavlov and Kasting, 2002; Zahnle *et al.*, 2006). Geologic $H_2S$ fluxes can be complemented by biological $H_2S$ fluxes originating from microbial sulfur reduction or sulfur disproportionation, common metabolic processes on Earth (Finster, 2008). The spectral properties of $S_8$ and $H_2SO_4$ aerosols differ with $S_8$ aerosols absorbing in the UV-blue region, while $H_2SO_4$ displays strong absorption at $\lambda > 2.7\,\mu m$. In principle, if volcanic $H_2S$ and $SO_2$ fluxes could be constrained, sulfur aerosol properties may indicate whether implied $H_2S$ fluxes imply an additional, biological source of $H_2S$, serving as a potential biosignature (Hu *et al.*, 2013). However, constraining volcanic sources remotely will be difficult and would require estimating the extent of subaerial versus submarine volcanism, which favors different $H_2S/SO_2$ outgassing proportions. More conservatively, characterizing sulfur haze properties would allow an independent assessment of the redox state of the atmosphere with $S_8$ indicating reducing conditions, and $H_2SO_4$ indicating oxidizing conditions. This could contribute to the overall appraisal of planetary habitability even if biogenic $H_2S$ fluxes were not constrained.

*4.2.8. Other gases.* The gases described above do not exhaust the list of volatile compounds produced by life on Earth, but encompass the unambiguous biogenic species widely believed to have been able to produce a measurable spectral impact at some point in Earth history. Other biogenic compounds are generated in abundance by Earth's biosphere but are not observed to rise to remotely detectable concentrations in a planetary disk average or are also produced in abundance by common abiotic processes. For example, isoprene is a common volatile organic compound produced by plants, phytoplankton, and animals, including humans (King *et al.*, 2010), but is quickly destroyed photochemically in Earth's oxic atmosphere (Palmer, 2003). Other secondary metabolic products (in contrast to direct products of metabolism such as $O_2$) fit this mold, and are reviewed in Seager *et al.* (2012; *e.g.*, their Table 3). Gases that are generated as products of metabolic processes, but also have common abiotic sources, encompass almost all simple molecules, including $H_2S$, $SO_2$, $N_2$, $H_2O$, $CO_2$, and many more.

## 4.3. "False positives" for biotic $O_2/O_3$ and possible spectral discriminators

The stated consensus as expressed in Des Marais *et al.* (2002) was that abiotic $O_2$ could be found on terrestrial exoplanets, but only on planets outside of the HZ (*i.e.*, either too close to the star or too far away to support habitable conditions). A tectonically active, water-rich planet with an active hydrological cycle was thought to have the capacity to remove abiotic $O_2$ from the atmosphere through geochemical or weathering reactions [*e.g.*, the reaction of $O_2$ with reducing volcanic gases or crustal ferric iron, Fe(II)].

The prevailing view has now evolved since Des Marais *et al.* (2002), with at least a few plausible mechanisms for generating $O_2$ abiotically on planets within the HZ, including from robust photolysis of $CO_2$ or a history of extreme hydrogen escape and $O_2$ buildup (*e.g.*, Domagal-Goldman and Meadows, 2010; Hu *et al.*, 2012; Domagal-Goldman *et al.*, 2014; Tian *et al.*, 2014; Wordsworth and Pierrehumbert, 2014; Gao *et al.*, 2015; Harman *et al.*, 2015; Luger and Barnes, 2015). For example, several authors have found that some abiotic $O_2$ (and $O_3$) could be produced in a prebiotic Earth-like, $N_2$-$CO_2$-$H_2O$ atmosphere with a surface ocean if the UV spectrum of the host star favored robust $CO_2$ photolysis (Domagal-Goldman *et al.*, 2014; Tian *et al.*, 2014; Harman *et al.*, 2015). These authors find that the extent of abiotic $O_2/O_3$ depends on a host of other boundary conditions (such as the flux of reducing gases and compounds), which may vary greatly from planet to planet. Significant hydrogen escape and $O_2$ buildup may be facilitated by atmospheres with low amounts of noncondensing gases lacking a cold trap (Wordsworth and Pierrehumbert, 2014) or an extended greenhouse phase after planetary formation due to the protracted superluminous premain sequence evolution of M dwarf stars (Luger and Barnes, 2015; Tian, 2015). These $O_2$ buildup scenarios need not completely deplete the entire $H_2O$ reservoir of the planet, but may exhaust $O_2$ sinks [such as crustal Fe(II)].

Each of the abiotic processes described above would generate their own spectral fingerprint, such as the simultaneous spectrally detectable presence of CO and $O_2$ from $CO_2$ photolysis (Harman *et al.*, 2015; Schwieterman *et al.*, 2016) or the absence of $N_2$ (Wordsworth and Pierrehumbert, 2014), which may be revealed through a lack of ($N_2$)$_2$ absorption (Schwieterman *et al.*, 2015b) or constraints on atmospheric mass through Rayleigh scattering (*e.g.*, Benneke and Seager, 2012). Highly evolved atmospheres that have experienced significant H-loss could be identified by highly density-dependent $O_4$ features (Schwieterman *et al.*, 2016) and by extended scale heights in transmission spectroscopy in the case of O-enriched He-dominated atmospheres (Hu *et al.*, 2015). Conversely, the existence of plausible mechanisms for abiotic $O_2$ strengthens the case for searching for biosignature couples such as $O_2 + CH_4$, possibly establishing more robust evidence for life. Other novel routes for abiotic $O_2$ buildup require highly desiccated atmospheres in tandem with the UV spectral energy distribution found in M dwarf stars (*e.g.*, Gao *et al.*, 2015), which may be identified through the absence of conventional habitability markers such as water vapor.

Considering all the above, our understanding of the plausible abiotic mechanisms for the presence of $O_2$ in exoplanet atmospheres is accompanied by action points that would assist in designing instruments and strategies for observing and characterizing potential biosignatures. This topic is explored extensively in Meadows (2017) and a companion article in this issue, Meadows *et al.* (2018).

## 4.4. Biosignatures in other types of atmospheres

It has been shown that atmospheres with $H_2$-dominated compositions could be habitable even with little $CO_2$ and at instellations lower than those predicted for the outer edge of the traditional HZ (Stevenson, 1999; Pierrehumbert and Gaidos, 2011; Seager, 2013). In addition, studies of exoplanet demographics from the Kepler mission have found that the most common type of planet is intermediate in radius (and mass) between Earth and Neptune (Batalha, 2014; although this conclusion is limited by the completeness of the Kepler sample, which is biased toward short-period planets). Referred to as "super-Earths," the stronger gravity of these planets increases the likelihood that their atmospheres could contain a low-molecular-weight component (i.e., $H_2$, He) over geologic time. (Note this does not suggest super-Earths may have a massive $H_2$ envelope like that of Neptune, which is not supported by data [e.g., Rogers, 2015]; the low-molecular-weight component would have to be small enough, relative to the planet's overall mass, to have a negligible impact on bulk density. This does rule out a significant $H_2$ fraction of a thin, terrestrial atmosphere, which is a negligible portion of a rocky planet's overall mass.)

However, we do not currently have examples of either abiotic or biological scenarios for rocky planets with significant $H_2$ fractions in our Solar System, although both early Mars and early Earth may have had a climatically significant $H_2$ component that requires further study (e.g., Tian et al., 2005; Ramirez et al., 2013; Wordsworth and Pierrehumbert, 2013). This situation introduces additional challenges for establishing biosignatures. However, if robust biosignatures are identified for exoplanets with a substantial component of $H_2$ or He, they will be significantly more detectable in transit spectroscopy because they will increase the scale height and therefore the transmission depths of spectral features (Miller-Ricci et al., 2009; Seager et al., 2013b; Ramirez and Kaltenegger, 2017).

Many reducing, H-bearing gases (e.g., $CH_4, C_2H_6, H_2S$) are problematic biosignatures in these atmospheres because abiotic equilibrium or kinetic processes could efficiently produce them. Just as in high-molecular-weight atmospheres, the most compelling signatures will be those that have equilibrium or kinetic barriers that prevent them from being easily generated abiotically.

Plausible biosignatures in $H_2$-dominated atmospheres could include $NH_3$, $CH_3Cl$, DMS, and $N_2O$ (Seager et al., 2013a, 2013b). Seager et al. (2013a, 2013b) proposed that $NH_3$ may be generated biologically in atmospheres dominated by $H_2$ and $N_2$ via the reaction $N_2 + 3H_2 \rightarrow 2NH_3$ since it is exothermic (energy yielding). Kinetic barriers prevent this reaction from occurring spontaneously at habitable temperatures. However, false positives include outgassed $NH_3$ from primordial material in the solid planet, exogenous delivery of cometary material, buildup from low UV emission from the host star, and chemical equilibrium reactions if temperature conditions are met in the subsurface or deep in a thick atmosphere. $NH_3$ is thus far the only biosignature candidate unique to $H_2$-dominated atmospheres (Seager et al., 2013a, 2013b). $NH_3$ has major absorption complexes near 2.0, 2.3, 3.0, 5.5–6.5, and 9–13 $\mu$m.

DMS, $CH_3Cl$, and $N_2O$ are possible biosignature gases for both $H_2$-dominated and Earth-like atmospheres (Seager et al., 2013a, 2013b; Sections 4.2.4 to 4.2.6 and references therein). $N_2O$ would be an intriguing biosignature on an $H_2$-dominated world, because it would have no abiotic sources. However, it would not be generated from energy-yielding metabolism (since producing it would be energetically unfavorable in an $H_2$ atmosphere)—although it could be an incidental by-product of metabolism in niche cases.

Additional biosignature gases in novel environmental contexts are currently being explored, but are highly speculative. An alternative, complementary, approach to empirically examining biosignatures gases is instead to begin with "all small molecules" that may be produced by (exo)life and subsequently filter those gases by their potential buildup in an atmosphere and their potential spectral detectability (Seager et al., 2016). Potential future directions in assessing novel biosignature gases are explored more deeply in the Walker et al. (2018) companion article in this issue.

## 4.5. Effects of the host star spectrum on photochemistry

In studies of planetary atmospheres generated by the fixed supply of gases (e.g., volcanic outgassing at a specified rate), changing the host star's type, and by extension of the wavelength distribution of light impinging on the planet, can dramatically alter the steady-state composition of the atmosphere. As mentioned in Sections 4.2.1–4.2.2 and 4.2.4–4.2.5, changes in the UV environment when transitioning from a G-type to an M-type star lead to increases in the concentrations of some biogenic gases, such as $N_2O$, $O_3$, and DMS (Segura et al., 2005; Rugheimer et al., 2015a). This possibility holds true for other gases as well, such as the abiotic generation of $O_2$ mentioned in Section 4.3. In addition, these changes can be driven by a host star's age, since a star's spectrum changes with time, potentially even before the star has evolved on to the main sequence (e.g., Ramirez and Kaltenegger, 2014; Luger and Barnes, 2015). Finally, the amount of radiation the planet receives represents one part of the equation when determining whether the planet has liquid surface water. Water vapor photolysis drives several catalytic cycles in the modern Earth's atmosphere and is expected to play a similar role in the atmospheres of planets around other stars. Taken together, the host star's spectrum has primary and secondary roles in determining the chemical composition of a planet's atmosphere.

One persistent consequence of the host star's spectrum is the planet's UV environment, as mentioned previously. Because UV photons are responsible for most of the photochemistry occurring in a planet's atmosphere, decreasing amounts of UV can lead to high concentrations of gases otherwise destroyed by UV photolysis, perhaps to unphysical levels if other potential sinks are not accounted for. For example, Zahnle et al. (2008) pointed out that a dense, cold, $CO_2$-dominated early martian atmosphere would be unstable to rapid and irreversible conversion to CO and $O_2$ (a "runaway"). A similar result was found by Gao et al. (2015) for Mars-like planets orbiting M dwarf host stars. In these cases, the cold surface temperatures limited water vapor photolysis, which in turn prevented the efficient catalytic recombination of CO and $O_2$. For the $O_2$ false-positive mechanism outlined in Section 4.3 by Harman et al. (2015), this same process occurs due to a lack of UV photons, rather than a lack of water vapor. On the other end of the spectrum, large amounts of UV radiation, in conjunction with higher stellar luminosities, can drive the photolysis (and ultimately the loss) of water vapor (Luger and Barnes, 2015).

Returning to low-UV environments, each potential biosignature gas mentioned in Sections 4.2.1–4.2.5 is subject to UV photolysis. Because of this, the same sort of ''runaway'' behavior has been noted for both $CH_4$ and $N_2O$ for planets orbiting UV-inactive late M stars (Rugheimer et al., 2015a). These cases persist even with large increases in the UV flux (Rugheimer et al., 2015a), suggesting that, at least for planets orbiting stars that are typically UV-quiet, we might expect higher concentrations of relevant biosignature gases, but as mentioned previously, lower UV fluxes also give rise to false-positive scenarios.

### 4.6. Impacts of flares and particle events on biosignature gases

In addition to a host star's time-averaged UV irradiation, stars can generate flares, charged particle events, and coronal mass ejections. The strength and frequency of these types of events vary with the star's size and age. For example, the young Sun was likely much more active than it is today, potentially for hundreds of millions of years (Güdel et al., 1997), and smaller stars (M4 and later) tend to be active for even longer periods of time, up to 8 Gyr (West et al., 2008). For terrestrial planets, these events can drive the photochemical modification of the atmosphere away from steady state, and, if strong flares and events occur frequently enough, could result in those modifications becoming the norm. If the planet in question is like prebiotic Earth, this situation could lead to the buildup of $N_2O$ from abiotic nitrogen oxidation (Airapetian et al., 2016), which may constitute a false positive. Alternatively, higher XUV fluxes driven by an increased number and intensity in flares may also drive more substantial atmospheric escape, which may render a terrestrial planet uninhabitable—in line with estimates for planets in high time-averaged UV environments (e.g., Luger and Barnes, 2015; Airapetian et al., 2017). In addition, planets around smaller stars may be subject to increased cosmic ray fluxes, which can modify the concentrations of key biosignature gases through the creation of $NO_x$ radicals (Grenfell et al., 2012).

Besides potentially producing false positives and desiccating planetary atmospheres, high-energy events can decrease the prevalence of other biosignature gases. An Earth-like planet struck by a single strong stellar flare would see only a small decrease in $O_3$ concentrations initially due to enhanced UV irradiation, but charged particles would produce a much larger decrease occurring over longer timescales, weeks to months (Segura et al., 2010). Subsequent flares within the flare recovery window would be expected to aggravate this effect but may be insufficient to completely remove the $O_3$ layer (Tabataba-Vakili et al., 2016). This flare activity may introduce detectable concentrations of other gases such as $HNO_3$, potentially allowing for the characterization of flaring trends beyond the observational window (Tabataba-Vakili et al., 2016), which provides further motivation for better constraints on the X-ray, UV, and charged particle environment for observed exoplanets.

### 5. Surface Biosignatures

Life may alter the spectrum of the surface of a planet through a variety of mechanisms, which include absorption and reflection of light by pigments in living organisms, scattering by the physical structures of organisms (including individual organisms and community architectures), degradation products of biological molecules, fluorescence of pigments, and bioluminescence. Each of these mechanisms may produce remotely detectable biosignatures; however, not all such biological spectral phenomena may be widespread enough to be detectable at the global scale on Earth, and not all are without abiotic mimics. Below, we catalog surface biological spectra arising from photosynthesis, other pigments, and reflectance features associated with cellular protection and ecological functions, and chiral biomolecules.

We summarize the spectral properties of known biomolecules on Earth, but note that only one surface spectral signature, the VRE, has been confirmed to produce a unique biological fingerprint on the disk-averaged spectrum of our planet. Other biomolecules that have been suggested in the literature thus far remain subjects for further research. In general, even on Earth, the spectra of such molecules can be subject to variation according to environmental conditions, physiological status, and species differences, tempering the likelihood of attaining diagnostic features that fingerprint exactly the molecules in question as is the case for gaseous absorption spectra. However, ''edge'' spectra are potentially powerful surface biosignatures that can occur throughout the visible and NIR (Hegde et al., 2015; Schwieterman et al., 2015a; Poch et al., 2017) and whose expression can depend on a host of factors, including intrinsic chemistry of major biomolecules, pigment acclimation relative to the environmental photon flux spectrum, and evolutionary contingencies. While we begin to explore the question of how these signatures may manifest on another planet, Walker et al. (2018) further address the challenges involved and explicitly outline fruitful avenues for future empirical and theoretical studies.

### 5.1. Photosynthesis

Since the review by Des Marais et al. (2002), OP remains the source of the most robust known planetary-scale biosignature: atmospheric $O_2$ and the surface reflectance spectrum of vegetation, the VRE. Anoxygenic photosynthesis, which likely evolved before the oxygenic variety, produces potentially diagnostic surface features, described below, but no robust gaseous signatures are yet known. Because photosynthesis leverages incident energy from the host star, it is the metabolic process with the greatest potential to affect the planetary environment, and is therefore essential to understand when making predictions of planetary-scale biosignatures. At our current state of understanding, it remains uncertain what different pigment colors might result from photosynthesis under different stellar irradiation and what other biogenic gases might result from photosynthesis adapted to other planets. The precise developmental pathway of photosynthesis, and specifically the evolutionary sequence of OP, is still an active area of inquiry (see Meadows et al., 2018, this issue), and scientific understanding is not yet sufficient to say whether the same or similar route of evolution would be followed on another planet. However, light energy use and storage in photosynthesis obey certain universal principles that also apply for solar energy from photovoltaic cells and must apply as well on exoplanets. The use of light energy to drive the movement of electrons is a remarkable feat of nature, involving a coordination of quantum dynamics and redox chemistry. In lieu of having a complete catalog of specific wavelengths to target for spectral features of surface pigments, understanding the fundamentals of photosynthesis, and the unknowns, should allow us to speculate on alternative expressions of photosynthesis on other planets.

Below we summarize basic principles of photosynthesis as well as important unknowns, the molecular nature and spectral features of light harvesting pigments, and how the VRE results. We then review the body of work to date that has delved into potential alternative photosynthetic biosignatures in other environments, and recent discoveries about photosynthesis that currently both challenge and enhance our understanding and ability to predict photosynthetic biosignatures on exoplanets.

### 5.1.1. Principles of photosynthesis.

All life involves oxidation/reduction (redox) reactions to transfer electrons from one chemical species to another. Whereas chemotrophs acquire free energy from redox gradients in the environment, photosynthesis utilizes light to generate its own redox gradients to perform biochemical reactions. Photosynthesis is the ultimate expression of life adapted to a star: stellar light is converted to chemical energy and used to drive biosynthesis of organic matter from $CO_2$. Thus, photosynthesis can drive redox cycles in the environment to sustain life beyond the lifetime of the free geochemical energy originally available. The basic processes of photosynthesis can be found in textbooks (*e.g.*, Whitmarsh and Govindjee, 1999; Blankenship, 2014) and are briefly summarized below. The general net equation is as follows:

$$CO_2 + 2H_2A + h\nu \rightarrow (CH_2O) + H_2O + 2A$$

$H_2A$ is a reductant that is oxidized biochemically (not photolyzed) to provide electrons for biochemical reactions. For OP, the $H_2A$ reductant is $H_2O$, and for anoxygenic photosynthesis, it may be $H_2S$, $H_2$, $Fe^{2+}$, or other reductants. $h\nu$ is the input photon energy, where h is Planck's constant, $\nu$ is the photon frequency $c/\lambda$, where c is the speed of light, and $\lambda$ is the wavelength of the photon wavelength. $CH_2O$ represents a reduced form of carbon in a sugar or carbohydrate in which energy is stored. The above equation is a net endergonic reaction; without the input of light energy, the back reaction is energetically favorable.

How is the light energy used and how does this determine a photosynthetic biosignature, pigment or gaseous? The above is a net equation that involves several separate steps in sequence, which follow these basic features for quantum harvesting of light energy, illustrated in the redox potential diagram of a generic photosynthetic ''reaction center'' in Fig. 8.

(1) The photoelectric effect (Einstein, 1905):

- Light harvesting: Certain materials, such as pigments or semiconductors, absorb photons across a spectral range, causing excitation of their electrons (Fig. 8A).

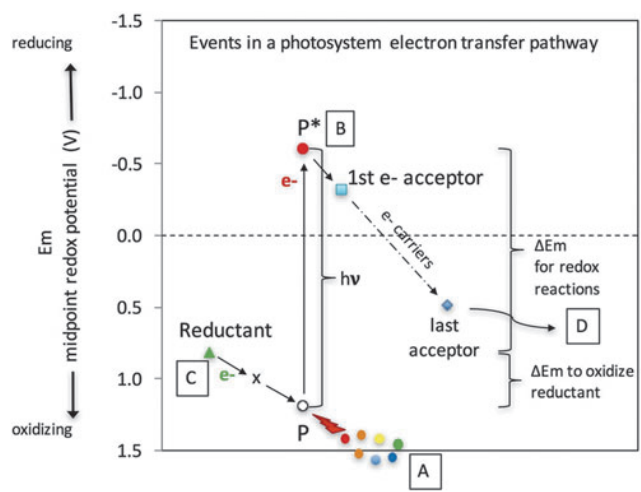

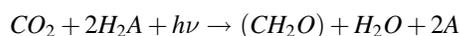

**FIG. 8.** Events in a photosynthetic reaction center electron transport pathway with noncyclic electron flow. Vertical axis shows redox potential in volts, with the convention being that negative is more reducing and positive is more oxidizing, with free energy reactions moving an electron from more negative to more positive. The P denotes the ground state of the primary photopigment, P* its excited state, and, not on the diagram, P+ is the state when an ejected electron has been captured, leaving an electron hole. **(A)** Antenna light harvesting pigments absorb photons across a spectral range and transmit the energy quanta to the primary photopigment at its ground state P, exciting an electron; the electron is excited to a reducing potential that is the difference of h$\nu$, the band gap, from the ground state potential P, creating an excited state P*. **(B)** The excited electron is ejected and quickly captured by the first electron acceptor to prevent it dropping back down to the ground state, creating an electron hole state P+, thus achieving charge separation. **(C)** The reductant is oxidized by the more oxidizing P+, and donates an electron to fill the hole; there may be an intermediate x along the way. **(D)** The first electron, having passed along a transfer pathway through various intermediate electron carriers, finally reduces a last electron acceptor for input to final energy storage products.

- Charge separation: At wavelengths of sufficient energy, shorter than the ''band gap*'' wavelength, the absorbing molecule ejects an electron, creating an electron hole, and an electron excited to a more reducing potential at a difference of hc/$\lambda$ from the ground state (Fig. 8B).

(2) Replacement of the ejected electrons: If producing electricity, closing the circuit will cycle back the electrons and generate current. In some types of anoxygenic photosynthetic bacteria, such cyclic electron flow is the primary mode of operation. In other phototrophs, noncyclic electron flow occurs, in which the originally ejected electron is instead captured by an electron acceptor molecule and then replaced via the primary photopigment oxidizing a reductant substrate, $H_2A$, which donates electrons (Fig. 8C). It is at this step that biogenic metabolic waste products from oxidation of the reductant (Equation product 2A) are produced.

---

*The band gap energy is the difference in energy between mixed orbitals where strong covalent bonds are formed. The lower energy filled orbital can have an electron excited into the higher energy orbital, and the energy gap corresponds to the photon energy required.

(3) Energy storage: The excited potential of the electron can be stored in a battery; in photosynthesis it is used for two things: to oxidize the reductant *and* in redox reactions along an electron transfer pathway to store the energy in reduced carbon from $CO_2$ via various carbon fixation pathways (Fig. 8D, and redox potential ranges indicated). Losses can occur to heat, fluorescence from de-excitation, and electrostatic discharge.

These steps must also occur for photosynthesis on an exoplanet. Perhaps the biochemistry will be different, but the same principles will apply. These basic features of light absorbance and energy use translate to biosignatures—pigment color and biogenic gases—through simultaneous environmental pressures and molecular energetic constraints.

*5.1.1.1. Relationship between band gap wavelength and reductant in the generation of biogenic gases and pigment color.* Because $O_2$ resulting from OP remains the most robust known biosignature gas, the question often arises how restricted is the suitable wavelength for enabling the extraction of electrons from water. Answering this question would clarify what exoplanet environments can support the use of which reductants for photosynthesis, and what pigment colors are likely to result. Referring again to Fig. 8, the band gap energy must straddle potentials that are both more oxidizing than the reductant (P more oxidizing than reductant at Fig. 8C), and more reducing than the first electron acceptor (P* more reducing than acceptor at Fig. 8B). A common misperception is that the photon energy all goes into direct photolysis of the reductant and that this explains the higher energy photons required for splitting the higher potential $H_2O$ compared to other reductants such as $H_2S$ or $H_2$. Instead the reductant is "split" via an oxidation reaction rather than by photolysis, and only part of the photon energy serves to drive that oxidation. Therefore, the ability to oxidize the reductant is solely due to the primary photopigment being more oxidizing than the reductant, not due to the photon energy input. While the redox potential of the reductant sets a strict oxidizing bound, the requirement of the first electron acceptor's redox potential may be a legacy of evolution with regard to how the overall photosystem structure arose, including protective mechanisms that have evolved, introducing inefficiencies (Rutherford *et al.*, 2012). Thus, there is no clear link between the band gap energy, and hence color, of the primary photopigment wavelength and the redox potential of the reductant.

*5.1.1.2. Uniqueness of OP.* The ability to extract electrons from water and produce $O_2$ as a waste product was transformative for our planet. Anoxygenic photosynthesis produces pigments that may serve as surface biosignatures, but the waste products from their reductants are not distinct from abiotic sources [*e.g.*, solid-phase Fe(III) from an aqueous Fe(II) substrate]. OP employs not one photosystem but two in series. The origin of this two-photosystem scheme, as well as the ability to oxidize water, is still unsettled and theories are reviewed in Meadows *et al.* (2018, this issue). Photosystem I (PSI) has its band gap at 700 nm and produces the final product for energy storage. Its ejected electron is replenished by photosystem II (PSII). PSII is responsible for extracting electrons from water, with a band gap at 680 nm. PSII is remarkable for having the most oxidizing biomolecule known in nature, sufficient to oxidize $H_2O$, a $Mn_4CaO_5$ cluster, the oxygen evolving complex (OEC). In Fig. 8, it occurs at location x at Step C. Until recently, the band gap wavelengths for PSI and PSII were the same for all known oxygenic phototrophs, from cyanobacteria to algae to higher plants, and thus, there was only one example of oxygenic photosystems on Earth. Recent discoveries of far-red oxygenic phototrophs now expand the sample space for extrapolating rules for exoplanets, which we detail in Section 5.1.4.

*5.1.2. Photosynthetic pigments and the color of phototrophs.* Light harvesting in photosynthesis is achieved through an array of different antenna pigments that can absorb across the whole VIS-NIR spectrum (Fig. 8A, colorful dots; Fig. 9). However, charge separation is only possible with a so-called reaction center (RC) pigment (Fig. 8, P and P*). This pigment can donate an electron after absorbing a quantum of light. The antenna pigments absorb at shorter wavelengths than the RC pigment and transmit that energy to the RC pigment, which traps the energy at its band gap wavelength. The relative spectral absorbance of both the antenna pigments and the RC pigment results in the color of the photosynthetic organism.

We now detail the special light-sensitive pigments used in photosynthesis. There are structural and chemical components of pigments that control the wavelengths of light that they absorb. It is useful to understand what these controls are because pigments may well be tuned to absorb at different wavelengths on exoplanets, and researchers should not necessarily expect to find the same absorption maxima, or VRE features.

*5.1.2.1. Structure.* In general, photosynthetic pigments are composed of a four-membered ring system called a tetrapyrrole. Addition of a fifth cyclopentene ring gives rise to a five-membered structure called the macrocycle. There are three types of pigment macrocycles: porphyrin, chlorin, and bacteriochlorin, which vary based on oxidation state or degree of saturation. These give rise to the major types of bacteriochlorophyll and chlorophyll pigments. Other structural components of the pigments include a central metal atom, peripheral substituents arrayed around the macrocycle, and a long hydrophobic esterifying alcohol (see Blankenship, 2014, and Allakhverdiev *et al.*, 2016, for molecular diagrams).

*5.1.2.2. Light absorption.* There are three main controls on the wavelengths of light that pigments absorb: (1) the oxidation state or degree of saturation of the macrocycle, (2) the functional groups arrayed around the periphery, and (3) interactions with surrounding proteins. A key point to note is that the absorption maxima of pigments are flexible and can be tuned by the phototroph based on environmental selection pressures.

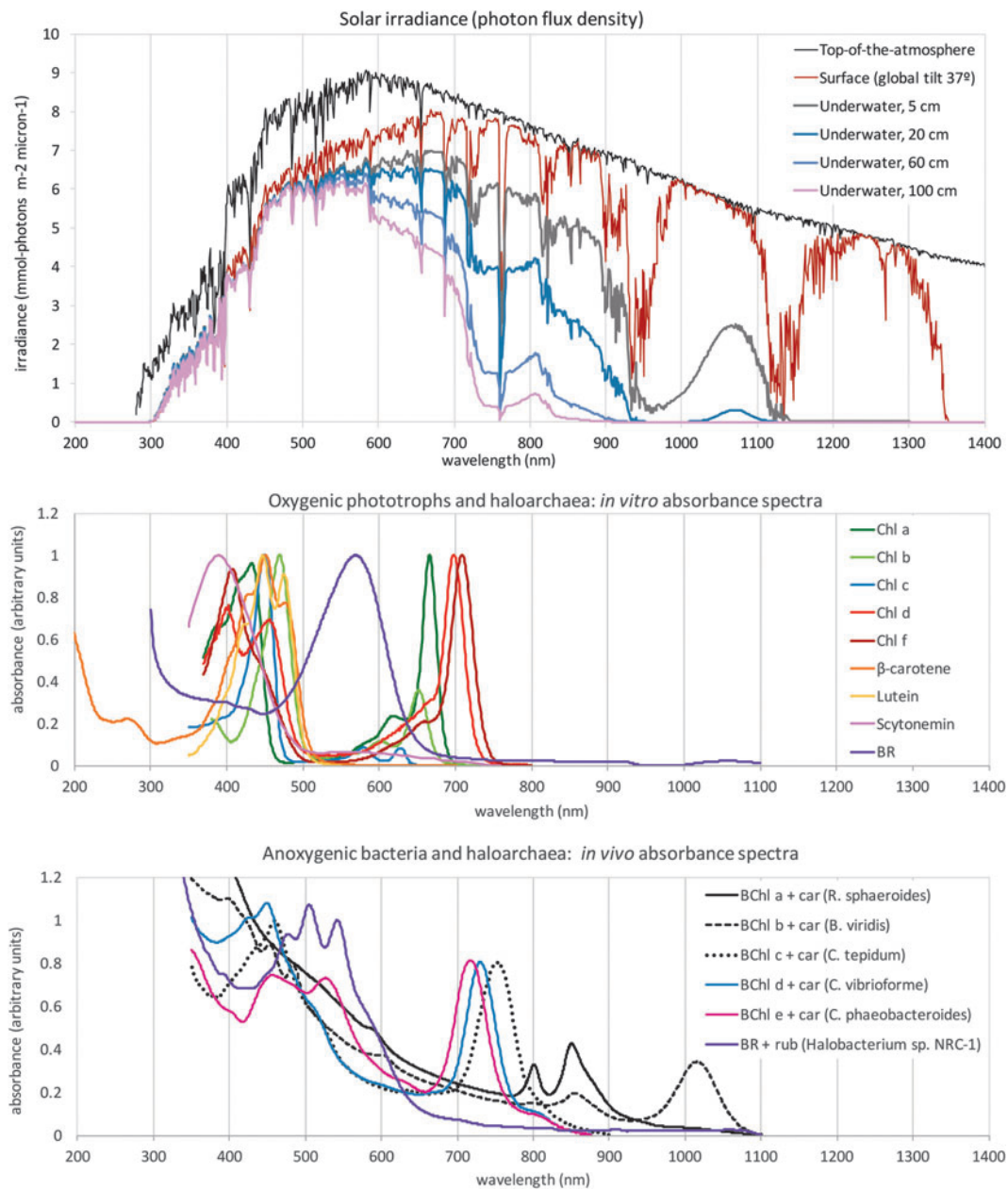

**FIG. 9.** Solar photon spectral irradiance on Earth and absorption spectra of pigments associated with phototrophic organisms, showing their peak absorbances at different wavelength niches. Pigment spectra are in arbitrary units scaled vertically for clear visualization of the wavelength of peak absorbance of the $Q_y$ band. Top: Solar spectral irradiance; top-of-the atmosphere and surface (National Renewable Energy Laboratory), and underwater at different depths (Kiang *et al.*, 2007b). Middle: Oxygenic phototrophic pigments *in vitro* (in solvent) absorbance spectra of chlorophylls (Chls), some carotenoids, the UV screening pigment scytonemin, and haloarchaea BR in purified membranes. Note that the *in vitro* spectra may be shifted to the blue by 5–40 nm compared with the *in vivo* spectra and should not be used for modeling activities. Bottom: Anoxygenic phototrophic pigments *in vivo* absorbance spectra of bacteriochlorophylls (Bchls) in whole cells, including their carotenoids, and a haloarchaeon containing bacterioruberins and BR (visible as shoulder, absorbance approximately 570–630 nm). Sources: Chl *a* in methanol, Chen and Blankenship (2011); Chl *b* in methanol, Chen and Blankenship (2011); Chl *c* in ether (Jeffrey, 1963); Chl *d* in methanol, Chen and Blankenship (2011); Chl *f* in methanol, Chen and Blankenship (2011), Chen *et al.* (2010), Li *et al.* (2012); beta-carotene in hexane, Dixon *et al.* (2005); lutein in solvent mixture, Harry Frank and Amy LaFountain, personal communication; scytonemin, Mueller *et al.* (2005); BR in sucrose gradient, Shiladitya DasSarma (coauthor), Priya DasSarma, Victoria J. Laye, personal communication, and Das-Sarma and DasSarma (2012); Bchl *a* and carotenoids in whole cells of *Rhodobacter sphaeroides* (purple anoxygenic bacteria), Richard Cogdell and Andrew Gall, personal communication; Bchl *b* and carotenoids in whole cells of *Blastochloris viridis* (purple anoxygenic phototroph), Richard Cogdell and Andrew Gall, personal communication; bacteriochlorophyll *c*, carotenoids, and minor content of Bchl *a* in whole cells of *Chlorobium tepidum* (green sulfur bacteria), Frigaard *et al.* (2002); Bchl *d* and carotenoids in whole cells of *Chlorobium vibrioforme* (green sulfur phototroph), Niels-Ulrik Frigaard, personal communication. All pigment spectra in this figure are available from the Virtual Laboratory Spectral Library Biological Pigments database (http://vplapps.astro.washington.edu/pigments). Bchls, bacteriochlorophylls; BR, bacteriorhodopsin; Chls, chlorophylls.

The six major pigments of anoxygenic phototrophs (bacteriochlorophylls) absorb in the NIR from ∼710 to 1040 nm (Table 1). The lower symmetry of the pigment structure relative to chlorophylls shifts the energy absorption bands to longer wavelengths. As a result, bacteriochlorophylls are well suited to absorbing the relatively higher flux of red and NIR radiation of M dwarf stars, the most abundant type of stars in our galaxy, as well as the plentiful flux of typical main sequence stars. The major bacteriochlorophylls are *a, b, c, d, e,* and *g* and are found distributed in Proteobacteria (purple sulfur bacteria, purple nonsulfur bacteria), Chlorobi (green sulfur bacteria), Chloroflexi (filamentous anoxygenic phototrophs), Firmicutes (heliobacteria), Acidobacteria (*Candidatus* Chloracidobacterium thermophilum) (Bryant *et al.*, 2007), and Gemmatimonadetes (*Gemmatimonas phototrophica*) (Zeng *et al.*, 2014).

The major chlorophyll pigments are chlorophylls *a, b, c, d,* and *f,* subsets of which are found variously in different types of oxygenic phototrophs—cyanobacteria, algae, and plants—all of which contain Chl *a* (see Kiang *et al.*, 2007a, for a breakdown, with update in Li and Chen, 2015). Anoxygenic green sulfur bacteria can also contain Chl *a*. Chl *a*, Chl *b*, and Chl *c* each have blue and red absorbance peaks slightly offset from each other. Chl *d* and Chl *f* are recent discoveries in cyanobacteria, which have the long wavelength absorbance peak shifted into the far-red/NIR.

Figure 9 shows absorbance spectra of the various (bacterio)chlorophylls, some carotenoids, and bacteriorhodopsin (Section 5.2) relative to the solar spectrum at Earth's surface and at various depths in water. These plots show how the various pigments absorb in different areas of the spectrum, which can reflect competition for light in the environment. The middle plot shows the *in vitro* absorbance spectra of pigments extracted in solvents, which shift the absorbance maxima blue-ward and do not reflect the true *in vivo* absorbance maxima of the pigments in living organisms. The bottom plot shows actual *in vivo* absorbance spectra, but note that even these spectra can be slightly shifted by ∼10 to 20 nm depending on the protein binding environment, which allows the phototrophs to tune their pigments to the available light.

Environmental pressures inevitably lead to tuning of light harvesting pigments to optimize the capture of photons. Compensation against excess light or other resource limits also occurs. Many organisms can acclimate their pigment mixes to fluctuating light environments in a behavior known as complementary chromatic acclimation (Kehoe and Gutu, 2006; Gutu and Kehoe, 2012). For example, plants can decrease their chlorophyll *a/b* ratio when shaded (Bailey *et al.*, 2001), and cyanobacteria can be triggered to synthesize chlorophyll *f* when exposed to far-red/NIR light (Gan *et al.*, 2014; Ho *et al.*, 2016).

The tuning of accessory pigments is constrained differently from the band gap wavelength of the RC pigment. The tuning of accessory pigments is likely to be matched to the peaks in the incoming light spectrum. Phototrophs can rapidly acclimate to fluctuating light conditions by synthesizing accessory pigments that absorb in specific areas of the spectrum. Thus, the tuning of these pigments for an exoplanet could be predictable given the spectral properties of the host star and radiative properties of the atmosphere. In contrast, the band gap of the RC pigments sets the upper bound on the range of wavelengths that can be used, and is the result of the legacy of evolution restricted by efficiencies achieved at the molecular scale. Therefore, the absorbance maxima of RC pigments may be unrelated or much more loosely coupled to the overall light environment of the planet, complicating predictive models. Anticipating the available mix of pigments and the wavelength of the primary donor pigment are areas for future research.

### 5.1.3. The vegetation "red edge".
Ultimately, for exoplanet biosignatures, we are interested in how these pigments manifest in the reflectance spectrum of their host organisms and the signal strength in a planet's radiance spectrum. The VRE is a well-known spectral reflectance signature of plant leaves. Des Marais *et al.* (2002) discussed the VRE as a surface biosignature of land vegetation, as it has been shown to be detectable in observations of Earth from the Galileo spacecraft (Sagan *et al.*, 1993), as well as in the Earthshine spectrum of light reflected from the Moon (Arnold *et al.*, 2002; Woolf *et al.*, 2002; Turnbull *et al.*, 2006). For surface signatures of photosynthesis, we can ask a series of questions. The primary questions are as follows: (1) can we expect to detect a red edge from vegetation (or cyanobacteria) on another planet, and at what other wavelengths could it be, (2) are there robust surface biosignatures for anoxygenic phototrophs, (3) what is the timescale and successional sequence of evolution of photosynthesis that would affect what might be observed in a time snapshot of the planet, (4) could non-photosynthetic pigments in chemotrophs likewise generate a detectable biosignatures, and (5) how do environmental factors like atmospheric opacity and surface mineralogy limit the detectability of a photosynthetic surface signal?

The VRE is so-called because of the strong contrast between absorbance in the red (660–700 nm) by Chl *a* (as well as Chl *b*) versus scattering in the NIR (∼760 to ∼1100 nm) due to the lack of absorbing pigment in this range and the change in the index of refraction between healthy mesophyll cell walls and air spaces in the leaf (Gausman *et al.*, 1974). It is fortuitous that in plants both the antenna (Ch *a* and Chl *b*) and band gap (Chl *a*) pigments share their peak absorbance in the red, matching the average peak in solar spectral photon flux. This contrast causes a steep, nearly step-like rise in reflectance from around 680 nm to a plateau at about 760 nm, spanning the red and far-red, as illustrated in Fig. 10—thus the name, "red edge." The strength of this contrast correlates with vegetation activity and quantity, since it is an index of both light absorbance for photosynthesis and the healthy, unstressed state of turgid cells in multiple layers of plant leaves. The precise midpoint of the red edge (or the derivative of the reflectance) can vary with physiological state and species. Nonetheless, it is ubiquitous among oxygenic photosynthetic organisms on land, and is so distinct from mineral signatures that it is regularly targeted with Earth observing satellites to identify the presence, activity, and type of vegetation on land (Ardanuy *et al.*, 1991; Friedl *et al.*, 2002).

TABLE 1. *In Vivo* Absorption Maxima of Light-Harvesting (Bacterio)Chlorophylls in Living Cells or Photosynthetic Membranes (After Pierson *et al.*, 1992)

| Pigment | In vivo *absorption maxima (nm)* |
|---|---|
| Bacteriochlorophyll *a* | 375, 590, 790–810, 830–920 |
| Bacteriochlorophyll *b*[a] | 400, 600–610, 835–850, 1015–1040 |
| Bacteriochlorophyll *c* | 325, 450–460, 740–755 |
| Bacteriochlorophyll *d* | 325, 450–460, 725–745 |
| Bacteriochlorophyll *e* | 345, 450–460, 710–725 |
| Bacteriochlorophyll *g* | 420, 575, 670, 788 |
| Chlorophyll *a*[b] | 435, 670–680 in PSII, 700 in PSI |
| Chlorophyll *b*[c] | 460, 650 |
| Chlorophyll *c*[d,e] | 442–452,[d] 580–587,[d] 630–632[e] |
| Chlorophyll *d*[f] | 710–720 in PSII, 740 in PSI |
| Chlorophyll *f*[g] | 720 |

[a]Brock Biology of Microorganisms; [b]Blankenship (2014); [c]Govindjee, personal communication; [d]Jeffrey (1969) in solvents; [e]Dierssen *et al.* (2015) in algae; [f]Mielke *et al.* (2013); [g]Gan *et al.* (2014).
PSI, photosystem I; PSII, photosystem II.

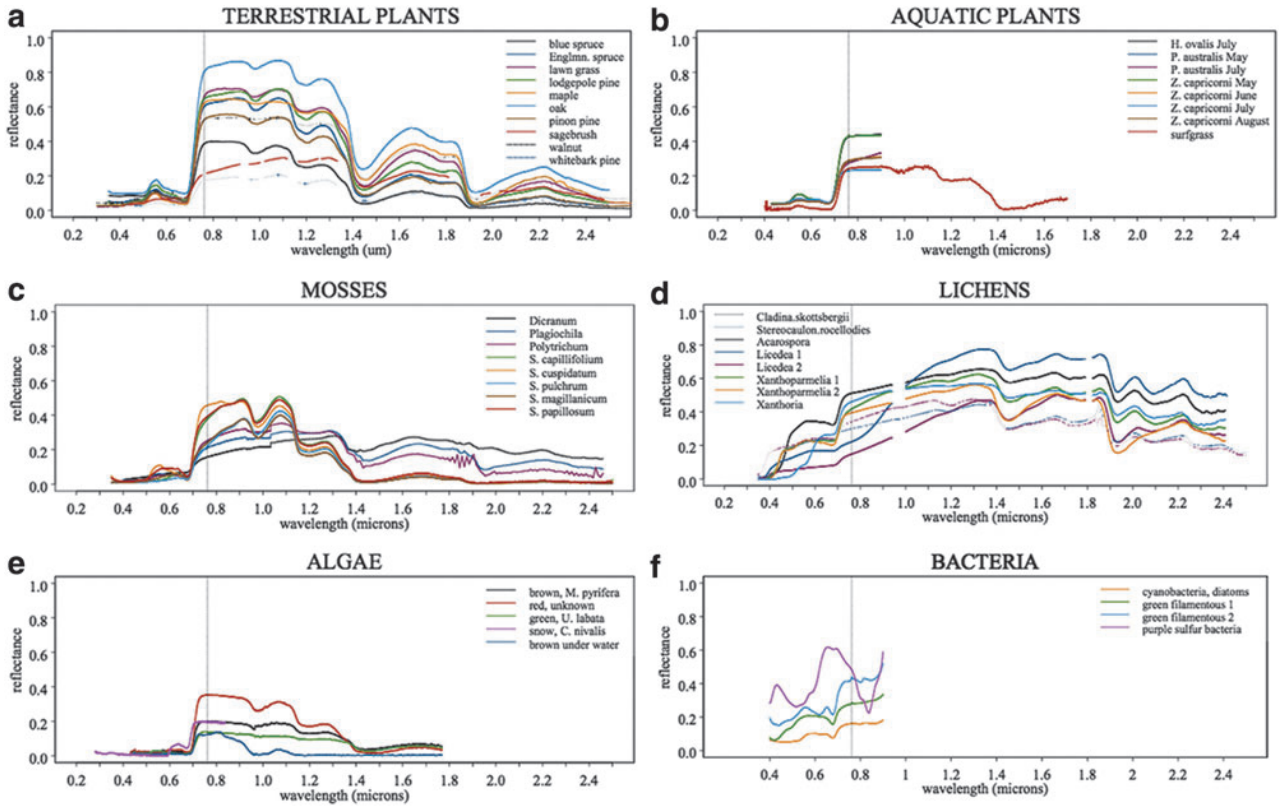

**FIG. 10.** The VRE occurs in all oxygenic photosynthetic organisms that use chlorophyll *a* **(a–e)** and cyanobacteria in **(f)**, with varying degrees of signal strength. Anoxygenic phototrophs use different kinds of bacteriochlorophylls absorbing in the near-infrared and thus general different reflectance features than the VRE. Source: Kiang *et al.* (2007a).

Given atmospheric corrections in Earth observations to obtain the surface reflectance, the most popular measure of the strength of the VRE is the Normalized Difference Vegetation Index (NDVI) = $(\rho_{NIR} - \rho_{Red})/(\rho_{NIR} + \rho_{Red})$, where $\rho_{NIR}$ and $\rho_{Red}$ are reflectance in the NIR and red, respectively (Huete *et al.*, 1994; Myneni *et al.*, 1995, 1997; Tucker *et al.*, 2005). The surface reflectance in the red would be in a narrow band centered around 650–680 nm where there is a peak in Chl *a* (and Chl *b*) absorbance. The reflectance in the NIR in plant leaves has a plateau that spans 760–1100 nm, but the satellite band used usually is cut off at <900 nm as sufficient to measure the NIR versus red contrast and avoid water bands that occur at the longer wavelengths. Some versions of this index may also use the visible absorbance in place of the red, since plants harvest light across the visible.

In disk-averaged spectra of a planet, atmospheric correction can be challenging. Arnold (2008) suggested a simple index, which he called the VRE index = $(\rho_{NIR} - \rho_{Red})/(\rho_{Red})$, and found its value in the Earthshine to be between 0% and 12%. This range is caused by variations in cloud cover, observing conditions, season, and Earth–Moon geometry. Arnold (2008) suggests 1% photometric precision would likely be required to detect an exact red edge analogue for a terrestrial planet with >10% vegetation coverage. Brandt and Spiegel (2014) used another fitting technique and similarly found that an SNR >100 would be required to detect the red edge on an Earth twin, assuming a spectral resolving power $(\lambda/\Delta\lambda)$ of R = 20. Tinetti *et al.* (2006a) estimated that detection of the red edge in a disk average would require at least 20% cloud-free diurnal average vegetated land cover. They note that because vascular plants promote conductance of water from the soil to the atmosphere and moisture convergence over land, cloud cover tends to favor vegetated land areas (classic example in Nair *et al.*, 2011), making detection of surface biota more challenging. Exploration of the coupled biosphere–atmosphere dynamics of clouds is an area for future work with exoplanet climate general circulation models (GCMs).

The red edge is much stronger than the "green bump" of chlorophyll *a* in the visible and has no (exact) abiotic mimics, and thus is a well-accepted biosignature (see Section 5.5). However, it remains an Earth-based biosignature, and so, its universality is still an open question, requiring exploration through the molecular mechanisms of photosynthesis summarized earlier. The next section reviews alternatives to the red edge that have been proposed.

### 5.1.4. Speculation about photosynthesis and pigment signatures on exoplanets.

Since Des Marais et al. (2002), the questions have naturally arisen as to whether the vegetation and cyanobacterial red edge could be adapted to be at some other wavelength on another planet, and whether other surface biological spectra besides those of OP could offer potential biosignatures. Searches for surface pigment signatures cannot always expect fixed wavelengths such as those found on Earth and must be interpreted within the environmental context of the star, atmosphere, and surface conditions of the planet.

Wolstencroft and Raven (2002) proposed that OP orbiting M dwarf stars could possibly use longer wavelength photons or three-photon systems in the NIR. Using a radiative transfer modeling approach, Tinetti et al. (2006b) showed that an "NIR edge" feature could potentially be more easily detectable than a red edge due to more favorable spectral positioning relative to water vapor absorption bands. Kiang et al. (2007a) conducted a comprehensive survey of types of photosynthetic organisms and their pigments, proposing that the wavelength of peak absorbance is likely to be adapted to match the wavelength of high photon flux density. Such correlation is consistent with the role of antenna pigments. With the VRE, it may be fortuitous that the band gap wavelength also occurs in the red. Other explorations of alternative light environments that might drive photosynthesis include studies for binary star systems (O'Malley-James et al., 2012), moons (Cockell et al., 2009c), and statistics of photosynthetically relevant HZ occurrence on planets in the Milky Way (von Bloh et al., 2010). Photosynthetic biosignatures at different stages of planetary evolution have been predicted. For example, an early Archean Earth surface biosignature from purple anoxygenic photosynthetic bacteria has been shown to be detectable in some simulated scenarios of possible coastal distributions (Sanromá et al., 2014).

While there are as yet no convincing theoretical alternatives to the tetrapyrrole-based pigments (Section 5.1.2) capable of energy transduction and charge separation (Mauzerall, 1976; Björn et al., 2009), various studies have investigated the efficiency of stellar spectral light absorption of these same pigments relative to other stellar types than the Sun (Komatsu et al., 2015; Ritchie et al., 2017), and therefore rationalize the potential for Earth organisms to survive on exoplanets based on light availability. However, because of the broad ability of pigment absorption maxima to be tuned via changes in the protein binding environment and changes in the chemical substituents in the pigment structure, as well as the ability of organisms to acclimate and adapt their mix of pigments, discerning photosynthetic pigment signatures around other stars requires deriving their plausible spectra within the given environmental context, not only from the available light spectrum but also from its intensity, its temporal variability, and other ecophysiological and ecological interactions that influence the ultimate expression of light harvesting pigments. "Edge spectra" of photosynthesis in other environments will be most likely due to accessory pigments, while the RC pigment will set the upper wavelength limit. Observed behaviors in phototrophs on Earth afford some level of predictability of pigment spectra, as exemplified by the discovery by Glaeser and Overmann (1999) of a previously unknown phototrophic purple bacterium by selectively enriching for a light niche not harvested by any known phototrophs at the time (in this case, the range was 920–950 nm). To improve confidence in such prediction for exoplanets may require Bayesian approaches to account for the unknown competing and consortia of organisms partitioning unidentified niches.

Some emerging work offers avenues for more confidently constraining surface signatures of both oxygenic and anoxygenic phototrophs. These include recent discoveries of far-red oxygenic phototrophs and observations of community-scale signatures of anoxygenic photosynthetic bacteria. Until recently, only one example of an oxygenic photosystem was known, with the pigment, chorophyll a (Chl a), as the primary electron donor pigment. The band gap wavelength for Chl a in PSII, at 680 nm, was long held to set the upper energetic limit for photons to oxidize water. The discovery of the far-red/NIR chlorophyll d (Miyashita et al., 1996) and chlorophyll f (Chen et al., 2010) both are both now challenging assumptions about the efficiency of OP (reviewed by Li and Chen, 2015). This newfound diversity in pigments for OP provides the opportunity now to uncover potential rules for alternative wavelengths for OP on other planets. Predicting the band gap wavelength and its long wavelength limit are subjects of theoretical research (see review in Walker et al., 2018, this issue), and the recent discoveries of far-red oxygenic phototrophs indicate that the long wavelength limit has not yet been reached.

Thus far, few biosignatures have been identified and accepted for an anoxic Archean-like planet. However, our planet was inhabited for very long periods before the evolution of land plants at ~0.5 Gyr (and the evolution of cyanobacteria <3.0 Gyr and the subsequent rise of $O_2$ at ~2.4 Gyr). A similar period of anoxic life may occur on exoplanets, and more studies are needed to characterize remotely detectable biosignatures associated with more evolutionarily ancient anoxygenic phototrophs. Recent studies have shown that anoxygenic phototrophs also produce edge-like features in the NIR due to bacteriochlorophylls in the range ~710 to 1040 nm (Kiang et al., 2007a; Sanromá et al., 2014; Schwieterman et al., 2015a; Parenteau et al., 2015; see Figs. 9 and 10). Anoxygenic phototrophs commonly occur in multilayered microbial mat communities. Parenteau et al. (2015) detected the major pigments in all layers of a mat from reflectance spectra at the surface due to their complementary spectral niches. This detection of multiple "NIR edges" on an exoplanet could signify layered phototrophic communities and possibly reduce the chance of a false positive due to mineral reflection (see section 5.4). Such "community biosignatures" should motivate more such explorations of community signatures in other ecosystems, such as marine intertidal areas, the open ocean, chemically stratified lakes and restricted marine basins, and other continental systems.

### 5.2. Retinal pigments

DasSarma (2007) proposed that the purple color of the retinal pigment in bacteriorhodopsin of haloarchaea could have served as a surface biosignature of an early Earth. Besides chlorophyll-dependent photosynthesis, utilization of light energy occurs also in many microorganisms that have evolved retinal-based pigments that enable utilization of light energy for ATP synthesis, phototaxis, vision, and other fundamental biological events (Ernst *et al.*, 2014). This light-driven proton pumping, however, is not considered photosynthesis because it is not connected to carbon fixation. The best-studied example is bacteriorhodopsin produced in the purple membrane of halophilic Archaea. Bacteriorhodopsin is the prototype of integral membrane proteins with seven-transmembrane α-helical segments bound to the retinal chromophore by a Schiff's base linkage to the ε-amino group of a lysine residue. Retinal is a $C_{20}$ lipophilic compound produced via the carotenoid biosynthetic pathway present across diverse phylogenetic groups and is simple enough to have evolved in the earliest cells on our planet. The light-driven proton pumping activity of bacteriorhodopsin can be coupled to ATP synthase to generate the energy currency in lipid vesicles, one of the simplest and potentially earliest bioenergetic mechanisms (Racker and Stoeckenius, 1974). Interestingly, retinal is a by-product of a major pathway leading to fatty acids, necessary for formation of lipid vesicle and cell membranes.

The central position of retinal at the intersection of lipid metabolism and bioenergetics, as well as its widespread distribution, suggests that this chromophore may have played an important role in the early evolution of life on Earth and possibly elsewhere in the Universe. In halophilic Archaea, bacteriorhodopsin imparts a bright purple color to cultures that can be observed through remote sensing (Dalton *et al.*, 2009). The strong color with an absorption maximum in the green region of the spectrum is complementary to the chlorophyll-based photosynthetic membranes (Fig. 9), and proposals for the coevolution of these pigments have been forwarded (Goldsworthy, 1987; DasSarma, 2006).

The bacteriorhodopsin-containing haloarchaea are aerobic heterotrophs, and therefore would have to have evolved after cyanobacterial OP. In addition, haloarchaea grow exclusively in hypersaline evaporitic settings, which may have been very limited in extent on the early Earth. However, recently a new genus of obligate anaerobic haloarchaea was discovered (Sorokin *et al.*, 2017), the lineage of which relative to its aerobic counterparts is yet to be established, but implying an expanding diversity of haloarchaeal metabolisms. The idea of an early Earth biosignature from retinal pigments calls for further research on the geochemical, geologic, and evolutionary context for potential biosignatures on exoplanets.

### 5.3. Alternative surface biosignatures: non-photosynthetic pigments and reflectance features

Organisms have developed pigmentation for a variety of purposes beyond light capture for carbon fixation or metabolic energy demands. The signatures of this pigmentation could also serve as viable surface biosignatures. The functions of pigmentation include screening of potentially damaging UV radiation (Proteau *et al.*, 1993; Solovchenko and Merzlyak, 2008; Archetti *et al.*, 2009), quenching of free radicals (Saito *et al.*, 1997; Cox and Battista, 2005; Tian *et al.*, 2008), protection against temperature extremes (Dadachova *et al.*, 2007; Liu and Nizet, 2009), scavenging of nutrients such as iron (Meyer, 2000), facilitating reactions between cells through the phenomenon of "quorum sensing" (McClean *et al.*, 1997; Williams *et al.*, 2007), protection against grazing through antimicrobial properties (Durán *et al.*, 2007), generating light through bioluminescence (Haddock *et al.*, 2010), and signaling of other organisms for purposes such as pollination (Chittka and Raine, 2006). This pigmentation may be completely decoupled from the light environment, the coloration incidental to function, as some microorganisms have been found to produce pigmentation in zero light conditions (Kimura *et al.*, 2003).

Schwieterman *et al.* (2015a) proposed that non-photosynthetic pigments could serve as alternative surface biosignatures and compiled a broad ranging classification of pigments involved in functions other than light harvesting, accompanied by optical spectral measurements of cultures of their host organisms (*e.g.*, Fig. 11, and publicly available through the VPL Spectral Database at http://vplapps.astro.washington.edu/pigments). In addition, Hegde *et al.* (2015) conducted an extensive survey of the reflectance signatures of extremophiles in a search for alternative potential biosignatures. The study by Hegde *et al.* (2015) included reflectance spectral measurements of pure cultures of 137 microorganisms with a spectral range of 0.35–2.5 μm, with results made publicly available (biosignatures.astro.cornell.edu). Generally, these studies showed that cultures of pigmented organisms contained the strongest signatures in the visible range, where pigment absorption was the strongest. Hegde *et al.* (2015) additionally found that bands of hydration (intracellular liquid water) at 0.95, 1.15, 1.45, and 1.92 μm appeared for all samples measured. These were the most significant signatures longward of pigment absorption in the visible range. Water bands of course are not unique to biology, but may indicate to the observer that the putative biological pigments are colocated with liquid water. This association with liquid water might be relevant for false positive discrimination. However, liquid water absorption bands may be difficult to disentangle from the water vapor absorption present in the planet's own atmosphere.

In contrast to photosynthetic pigments, which may have faced evolutionary pressure to adjust the wavelength of photon absorption based on their environment and energy needs, pigments adapted to other functions may produce "edges" throughout the visible wavelength range. This provides both an opportunity and a challenge for the identification of surface biosignatures. Non-photosynthetic pigments provide a larger Universe from which to search for edge signatures, but the wavelength at which that edge occurs may be uncorrelated with the stellar spectrum and therefore may be more difficult to interpret (see section 5.4 below for a deeper discussion on false positives for surface biosignatures).

The salt ponds and hypersaline lakes whose coloration is dominated by haloarchaea are examples of surface environments exhibiting alternative biosignature features. In these cases, pink, red, or orange coloration results from a proliferation of salt-tolerant archaea such as *Halobacterium salinarum* and bacteria such as *Salinbacter ruber*, whose carotenoid and rhodopsin pigments allow adaptation to the environment and dominate the coloration of the ponds (Oren *et al.*, 1992; Oren and Dubinsky, 1994). Figure 12 demonstrates the contrasts between potential spectral signatures resulting from red edge production vegetation (conifers) and halophile-dominated salt ponds such as those found in San Francisco Bay (Schwieterman *et al.*, 2015a). The 100% surface coverage for each type is unrealistic, but is intended only to demonstrate maximum differences in reflectivity rather than a reasonable expectation for the strength of a signature detected on an exoplanet. More understanding of the potential coverage of such environments on other planets is needed to determine what expectation is reasonable. In the far future, predictable spatial distributions of pigments (continental margin, latitudinal, *etc.*) could be compelling on an exoplanet.

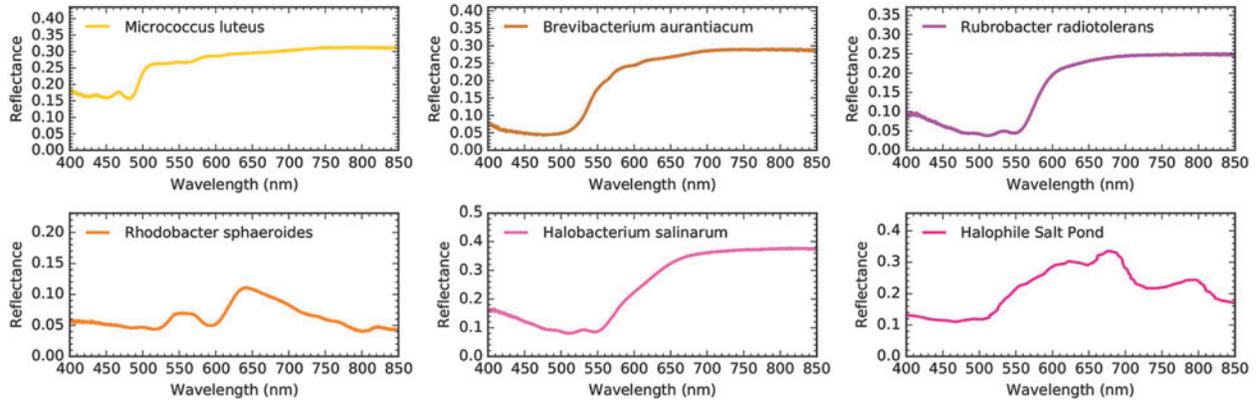

**FIG. 11.** Alternative surface reflectance biosignatures. Panels show reflectance spectra of a collection of non-photosynthetic organisms, oxygenic phototrophs, and anoxygenic phototrophs from Schwieterman *et al.* (2015a), except for the last panel, which is an environmental spectrum of a San Francisco saltern pond from Dalton *et al.* (2009) dominated by pigmented halophiles.

### 5.4. False positive surface biosignatures

Spectral edges produced by pigment-bearing organisms are potentially compelling surface biosignatures. The question naturally arises regarding how well this phenomenon uniquely discriminates biological from abiotic material. While the VRE itself has no exact spectral mimic among common abiotic materials, this is not the case for edge features in general. For example, Seager *et al.* (2005) propose mineral semiconductors as potential false positives for blue-shifted VRE analogues. The electronic band-gap energies in cinnabar and sulfur, for example, produce abrupt albedo increases at 0.6 and 0.45 µm, respectively (Fig. 13). Moreover, we can find examples within the planetary bodies of the Solar system that show this effect. Figure 13 shows the surface reflectance spectrum of Jupiter's moon Io, which shows a somewhat steep spectral increase between 0.4 and 0.5 µm (Karkoschka, 1994). This edge-like slope is generated from the sulfur compounds that coat Io's surface because of vigorous tidally induced volcanic eruptions.

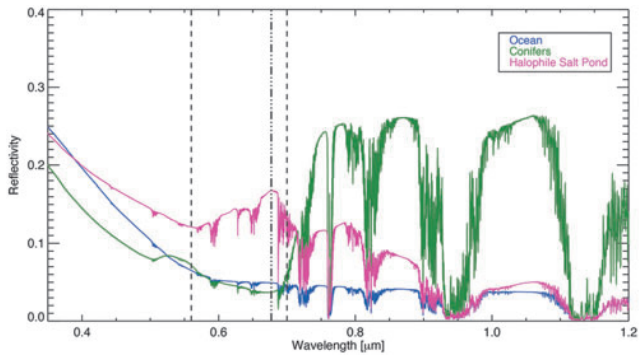

**FIG. 12.** Reflectivity contrasts for surface biosignatures. Synthetic direct-imaging spectra of model planets with an Earth atmosphere whose surfaces are dominated by an ocean (blue), a conifer forest (green), and a halophile-dominated saltern pond (pink). The dashed line represents the spectral slope of the pure *Halobacterium salinarum* culture shown in Fig. 13. The dot-dash line denotes the 0.68 µm spectral peak suggested as a halophile reflectance signature. The conifer and salt pond spectra were generated using reflectance spectra sourced from Baldridge *et al.* (2009) and Dalton *et al.* (2009), respectively. This figure is reproduced from Schwieterman *et al.* (2015a) with permission from the authors who retain the copyright.

This false positive issue is particularly acute for low-resolution spectrophotometric measurements that may attempt to quantify spectral transitions in a similar way to the NDVI. In addition to the problem noted for mineral semiconductors above, many more surfaces within the Solar system exhibit sloped increases from red to infrared wavelengths, although more gently sloped than the VRE. An example of this is described by Livengood *et al.* (2011), who find that the NDVI for the Moon is substantially greater than for Earth's disk-averaged spectrum.

### 5.5. Chiral and polarization biosignatures

Chiral compounds such as amino acids, sugars, and nucleic acids are asymmetric molecules whose mirror images cannot be superimposed on one another. During abiotic synthesis of these compounds, equal proportions of each "handedness" or enantiomer are produced, generating a racemic mixture. Organisms display a preference for building larger molecules out of one enantiomeric form. For example, all known organisms, including bacteria, archaea, eukaryotes, and even viruses, encode left-handed amino acids into proteins and right-handed sugars into multiple biopolymers. It is currently unknown how the biological preference for certain enantiomers arose from racemic mixtures in a prebiotic world, and it is an area of intense research. However, given this biological preference, enantiomeric excess of these compounds is considered a powerful biosignature. Chirality could be generic to all life, and hence, this biosignature has the potential to reveal extant organisms even if they differ substantially from terrestrial life.

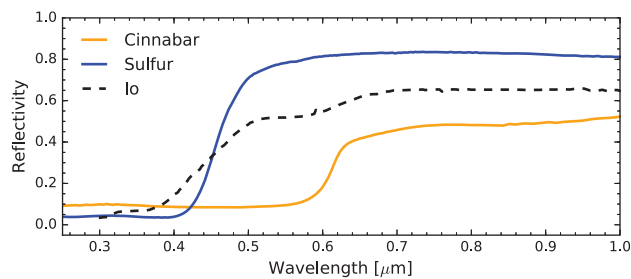

**FIG. 13.** Potential ''false positives'' for spectral ''edge'' biosignatures. The reflectance spectra of elemental sulfur (blue) and cinnabar (orange) are sourced from the USGS spectral library (Clark *et al.*, 2007). The reflectance spectrum of Io is sourced from Karkoschka (1994).

There are some notable exceptions to abiotically synthesized racemic mixtures. Slight enantiomeric excess (<10%) of a few amino acids has been detected in some carbonaceous meteorites (Elsila *et al.*, 2016 and references therein). However, these amino acids are either rare or undetected in Earth's biosphere. An enantiomeric excess of sugar acids was recently found in a variety of carbonaceous meteorites (Cooper and Rios, 2016). However, in general, enantiomeric excess greater than ∼20% can be considered a robust biosignature.

Amino acids and sugars are optically active in the UV, and the chiral light-absorbing centers preferentially absorb left-handed or right-handed circularly polarized light, giving rise to features in circular polarization spectra. In fact, biochemists have been using circular polarization, or circular dichroism, for 30+ years to characterize the secondary structure of proteins (Kelly and Price, 2000). Photosynthetic pigments are also optically active in the VIS-NIR and contain chiral centers. When these pigments are present in aggregates displaying long-range chiral order, they give rise to the sine curve-looking feature in the circular polarization spectrum (Garab and van Amerongen, 2009; Sparks *et al.*, 2009a, 2009b; Patty *et al.*, 2017) (Fig. 14). This feature corresponds to the absorbance maxima of the pigment. These ''psi-type'' (polymer and salt-induced) spectral features are much stronger than features from isolated pigments that contain individual chiral centers.

Chirality can in principle be remotely detected on planetary scales by linear and circular polarization spectroscopy. Linear polarization (light waves that oscillate in single direction rather than rotating in a plane as the light travels) has been used in biochemical studies to characterize, for example, the orientation of pigment molecules in cells, while circular polarization has been used to examine excitonic coupling in pigment–protein complexes (Garab and van Amerongen, 2009). Investigators have examined whether pigment signatures in linear polarization spectra could serve as remotely detectable signs of life (Berdyugina *et al.*, 2016). The degree of linear polarization is correlated with the strength of pigment absorption, producing an inverted effect relative to the VRE in which the degree of linear polarization is high at visible wavelengths and low in the NIR (Peltoniemi *et al.*, 2015; Berdyugina *et al.*, 2016). However, it has been shown that abiotic environmental factors such as mineral dust (West *et al.*, 1997), scattering by particles in the atmosphere (Shkuratov *et al.*, 2006), and the molecular absorption by gases (Stam, 2008; Takahashi *et al.*, 2013; Miles-Páez *et al.*, 2014) can also produce positive linear polarization signals. As a result, the biotic and abiotic inputs of linear polarization spectra must be disentangled. In contrast, the degree of circular polarization is more directly tied to the optical activity of amino acids and pigments, making it a more specific biosignature (Sparks *et al.*, 2009a, 2009b; Patty *et al.*, 2017). However, of concern is the relatively weak signal strength of the circular polarization signal of photosynthetic pigments, which is on the scale of $10^{-2}$–$10^{-4}$ (Sparks *et al.*, 2009a, 2009b). In addition, the degree of polarization responds to the physiological state of the organisms, potentially adding a predictable temporal variation to the polarization signature (Patty *et al.*, 2017).

Sterzik *et al.* (2012) measured the linear polarization of Earthshine and found that for one observation window, the data were best fit with a model containing 10–15% noncloud covered vegetation. A circular polarization signal was not detected. However, it is not known whether light reflecting from the dusty surface of the Moon interfered with the polarization signal. Linear and circular polarization signatures providing an additional, potentially determinative, dimension for biosignature analysis are interesting topics for future study.

### 5.6. Fluorescence and bioluminescence

Photons directly sourced from organisms could represent another category of surface biosignature. One manifestation of this phenomenon is chlorophyll autofluorescence, which has been observed by low Earth-orbiting satellites with the aim of characterizing plant health and primary productivity (Joiner *et al.*, 2011). Autofluorescence is the reprocessing of absorbed higher energy photons into emitted lower energy photons. Autofluorescence is observed in abiotic materials such as fluorite and calcite, so biogenicity must be determined from the nature of the fluorescence spectrum. Chlorophyll fluorescence is a signature of photosynthesis believed to reduce physiochemical stress on the organism (Papageorgiou and Govindjee, 2007). The fluorescence spectrum of Chlorophyll *a* consists of broad emission from 640 to 800 nm with maxima located at 685 and 740 nm (Meroni *et al.*, 2009). Because fluorescence is the instantaneous reprocessing of absorbed light, it occurs only on the day side of the planet, and thus, the fluorescence signature must be measured against reflected light from the Sun [or star(s)]. The chlorophyll fluorescence signature over vegetated areas constitutes about 1–5% of the total spectral flux at the fluorescing wavelengths (Meroni *et al.*, 2009; Joiner *et al.*, 2011). To detect the fluorescence component amid contamination from reflected light, observations focus on telluric (*e.g.*, the $O_2$-A band) or reflected solar absorption bands (*e.g.*, H-alpha), which are partially filled in by the fluorescent emission flux. The fluorescence flux constitutes a proportionally larger fraction of the total flux at these wavelengths (Meroni *et al.*, 2009; Joiner *et al.*, 2011). Fluorescence as an adaptation in response to high-UV flare events has been proposed as a potentially temporal biosignature for planets orbiting M dwarf stars (O'Malley-James and Kaltenegger, 2016). To detect biological fluorescence will be very challenging and would require signal-to-noise ratios higher than for most features thus far discussed.

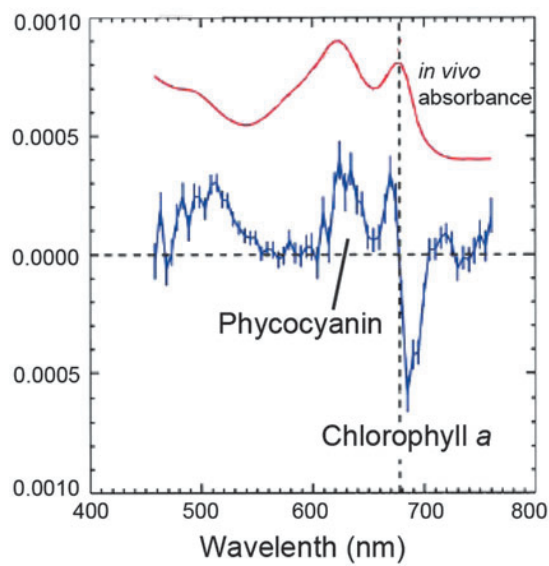

**FIG. 14.** Reflected circular polarization spectrum of a planktonic suspension of a marine cyanobacterium (center plot). The *in vivo* absorption spectrum is shown in the upper plot and is arbitrarily scaled. The absorbance maxima of Chl *a* (435, 680 nm), phycocyanin (620 nm), and carotenoids (∼450 nm) are visible. The large feature in the circular polarization spectrum corresponds to the absorbance by chiral aggregates of Chl *a* and proteins. After Sparks *et al.* (2009b).

Bioluminescence, in contrast to fluorescence, involves the direct production of photons through the oxidation of a luciferin ("light-bringing") molecule (Haddock *et al.*, 2010). Luciferin is a general category of light-emitting molecules involved in bioluminescence, produced by a wide variety of lineages that independently evolved bioluminescence, including species of bacteria, fish, eukaryotic planktons, and insects (Haddock *et al.*, 2010). Most luciferin molecules have a peak emission wavelength in the green region of the visible spectrum (in contrast to chlorophyll fluorescence that peaks in the red). *Vibrio* bacteria in the ocean can generate a faint bioluminescent glow covering up to 10,000–20,000 km². Called a "Milky Sea" effect, this glow can be characterized by Earth observing satellites (Miller *et al.*, 2005). The potential of bioluminescence to serve as an exoplanet biosignature is poorly studied and constrained. However, it is important to point out that bioluminescence will be limited by the productivity and ecology of the biosphere and more stringently on its evolutionary function, which may vary but will likely be very challenging to detect on exoplanets due to signal-to-noise constraints.

## 6. Temporal Biosignatures

Temporal biosignatures are measurable time-dependent modulations that indicate the presence of a biosphere acting on a planetary environment (*e.g.*, Meadows, 2006, 2008). These temporal modulations can take the form of oscillations in gas concentrations or the surface spectral albedo of the planet. They may even cause the direct emission of light by organisms (*e.g.*, bioluminescence, see Section 5.6) provided there is a direct link between the detectable change (*e.g.*, stellar activity) and biological action. The most commonly referenced temporal signature is the seasonal change in $CO_2$ concentration in Earth's northern hemisphere as a response to the changing productivity of the land biosphere as a function of temperature and insolation (*e.g.*, Keeling *et al.*, 1976). Seasonal changes may be driven primarily by obliquity, as on Earth, or by orbital eccentricity where the effects from changing planet–star distance would dominate. However, a temporal biosignature need not be seasonal in cadence but could be diurnal or act in response to another measurable environmental variable. Temporal biosignatures have been less studied than other types of biosignatures, in part, because of the complexity required to model them with additional variables such as the axial tilt, orbital eccentricity, and surface heterogeneity of the planet likely playing key roles. In addition, it is important to note that temporal changes must be present in the disk average to be observable and therefore will be dependent on observer/target viewing geometries in addition to the static properties and intrinsic changes of the planetary environment, particularly for those seasonal changes driven by obliquity. For example, an equator-on view will average changes in both visible hemispheres and require hemispherical asymmetries for a seasonal change driven by obliquity to be observed. However, in the other extreme, a near pole-on view will primarily include only one hemisphere, and so a hemispherical dichotomy would not be necessary to observe seasonal changes. Most scenarios will reside between these extremes. Of course, temporal changes driven by eccentricity would also not require specific hemispherical dichotomies, although they may complicate interpretations should they exist. In this section, we present an overview of temporal biosignatures based primarily on deriving inferences from Earth system studies and observations.

### 6.1. Oscillations of atmospheric gases

Earth's biosphere imparts modulations on several key spectrally active gases in the atmosphere, including $CO_2$, $O_2$, $O_3$, and $CH_4$ (Fig. 15). The most well known of these is the seasonal oscillation in $CO_2$ due to growth and decay of vegetation on land (Keeling, 1960; Hall *et al.*, 1975; Keeling *et al.*, 1976). The $CO_2$ content of the atmospheres decreases in the spring season as $CO_2$ is fixed into organic matter by vegetative growth, then begins to rise in fall and winter as $CO_2$ consumption slows and plant matter decays. The magnitude of the change is dependent on hemisphere and latitude. In the northern hemisphere, the amplitude of the $CO_2$ oscillation ranges from $\sim 3$ ppm near the equator to $\sim 10$–20 ppm at high latitudes (Keeling *et al.*, 1996). Overall, the amplitude is far greater in the northern hemisphere than the southern due to its substantially greater continental area (and thus vegetative cover). The seasonal variation in $O_2$ concentrations is linked to $CO_2$ consumption by the photosynthesis net reaction ($CO_2 + H_2O \rightarrow CH_2O + O_2$) and the reverse decomposition net reaction ($CH_2O + O_2 \rightarrow CO_2 + H_2O$), so that oscillations in $O_2$ are anticorrelated with those of $CO_2$. The absolute amplitude of $O_2$ variability (in ppm or per moles), however, is larger than that of $CO_2$ because $CO_2$ is substantially more soluble in seawater (Keeling and Shertz, 1992). The lower solubility of $O_2$ also explains why its magnitude of seasonal variability is comparable between hemispheres, with an amplitude of about $\sim 50$ ppm at midlatitudes (Keeling *et al.*, 1998).

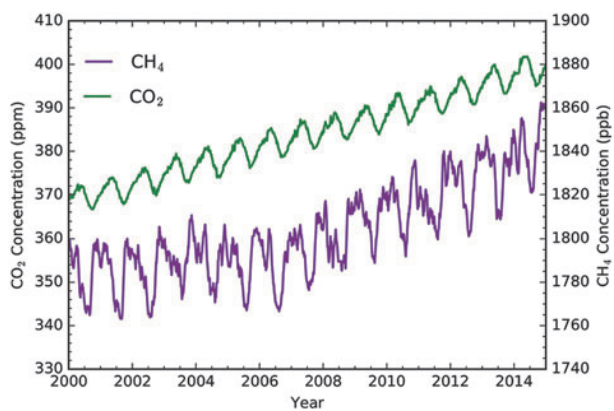

**FIG. 15.** Gas abundance oscillations as a possible temporal biosignature. Volume mixing ratio measurements of $CH_4$ and $CO_2$ from the National Oceanic and Atmospheric Association (NOAA) at Mauna Loa, HI from 2000 to 2015 (Dlugokencky *et al.*, 2017; Thoning *et al.*, 2015). The seasonal variations in both gases are partially reflective of the seasonal change in the productivity of the biosphere in the northern hemisphere (the secular increase in both gases is attributable to industrial emissions). These data were obtained from the NOAA's Earth System Research Laboratory (www.esrl.noaa.gov).

Seasonal variability in $CH_4$ is more complicated. The highest $CH_4$ levels occur in the late fall and early spring, but $CH_4$ concentrations fall in the summer and winter, reaching an annual minimum in northern summer and a more muted minimum in winter (Rasmussen and Khalil, 1981). The variation in $CH_4$ concentrations is only partly biogenic (*i.e.*, controlled by biological sources and sinks); instead, its temporal oscillations are dominantly controlled by interactions with OH ions that destroy it ($OH + CH_4 \rightarrow CH_3 + H_2O$). This destructive OH is ultimately sourced from tropospheric $H_2O$ [mostly via $H_2O + O(^1D) \rightarrow 2OH$], which increases in summer (due to increased water evaporation) and decreases in winter, so that the $CH_4$ abundance in the atmosphere is more strongly correlated with surface temperature (hence $H_2O$) and solar zenith angle [hence $O(^1D)$] than instantaneous release from the biosphere (Khalil and Rasmussen, 1983). This example illuminates an interesting scenario through which a gas is primarily sourced from biological activity, but its seasonal modulation is abiotic. $O_3$ on Earth also displays a dominantly abiotically driven seasonal variability with a midlatitude early spring maximum due to suppressed photolytic destruction in winter, which has an amplitude of about 30% in the total column value (WMO, 1994). Temporal variation in the $O_3$ column can arise due to both photochemical and dynamical effects.

Detecting gas oscillations of the same magnitude as those of Earth's modern biosphere will be quite challenging and likely beyond the capability of next-generation observatories. $CO_2$ and $CH_4$ vary on the order of 1–3% (and $CH_4$'s variation is only partly biogenic), while $O_2$ varies by 50 ppm against a 21% background, or by $\sim 0.02\%$. In addition, target absorption bands (such as $CO_2$'s 15 µm band) will not vary linearly with abundance if they are saturated; thus, the measurable spectral variability may be less than this. In ideal circumstances, temporal gas oscillations will occur at background gas abundances that are detectable, but not saturated, and impart measurable spectral variability. In addition, potential false positives, such as seasonal sublimation of $CO_2$ ice, must be ruled out but could be in cases where planetary temperature ranges are known (*i.e.*, as seen on Mars). Further work is required to illuminate the cases in which a biosphere could produce measurable and inferable modulations of gases linked to metabolism.

### 6.2. Oscillations in surface signatures

Changes in surface albedo or the spectrum of reflected light from the surface represents another form of temporal biosignature. For example, the VRE signature is temporally variable because of the seasonal growth and senescence of green vascular plants on continents (Fig. 16). The temporal variability of the "edge" signature, in conjunction with seasonal changes at the same phase, could enhance the case for the biological origin of the signature. Furthermore, the NDVI of stressed and dead vegetation has a lower value than living vegetation, because chlorophyll absorption is weaker (making the plant more reflective in the visible), while desiccated vegetation reduces the NIR reflectivity (Tucker, 1979). The circular polarization fingerprint of vegetation also changes as a function of physiological stress (Patty *et al.*, 2017) and may be an additional temporal signature in phase with reflectivity changes. Changing external conditions may also induce changes in pigmentation to acclimate the organisms to temperature or irradiation stressors (*e.g.*, Archetti *et al.*, 2009). It is important to note that microbial mats also experience seasonal alterations in pigmentation (Nicholson *et al.*, 1987). Time-dependent changes in bioluminescent or biological fluorescence could also serve as a temporally variable surface biosignature (Section 6.3). Necessarily, seasonal changes in the reflection signature will be smaller than or equal to the maximal steady-state spectral signature, so greater signal-to-noise will be required to measure them. As stated previously, temporal biosignatures need not necessarily be driven by seasonal changes, but could also occur in phase with rotation or other dynamical processes. For example, tidally induced algal blooms have been posited as a potential remote temporal biosignature (Lingam and Loeb, 2017).

### 7. Assessing Biosignature Plausibility

Methods for assessing biosignature plausibility are important for two primary reasons. First, there is the problem of interpretation. Because of potential false positives, the detection of a single gas, and perhaps even some combination of gases, is unlikely to constitute robust evidence of life alone. Therefore, whether we attribute measured spectral properties of an exoplanet to the presence of life requires a scheme for evaluating the planetary context to determine the likelihood that a biological process is involved. Second, we wish to prospectively determine the best biosignatures to search for as we design the technological capabilities to characterize exoplanets for evidence life. This goal is essential, as different gases and surface features have spectral features at different and often widely separated spectral locations that will be impossible for any single instrumental configuration to span. Below, we briefly review three potential methods for evaluating biosignature plausibility: chemical disequilibrium, biomass estimation, and applications of network theory to biosignatures.

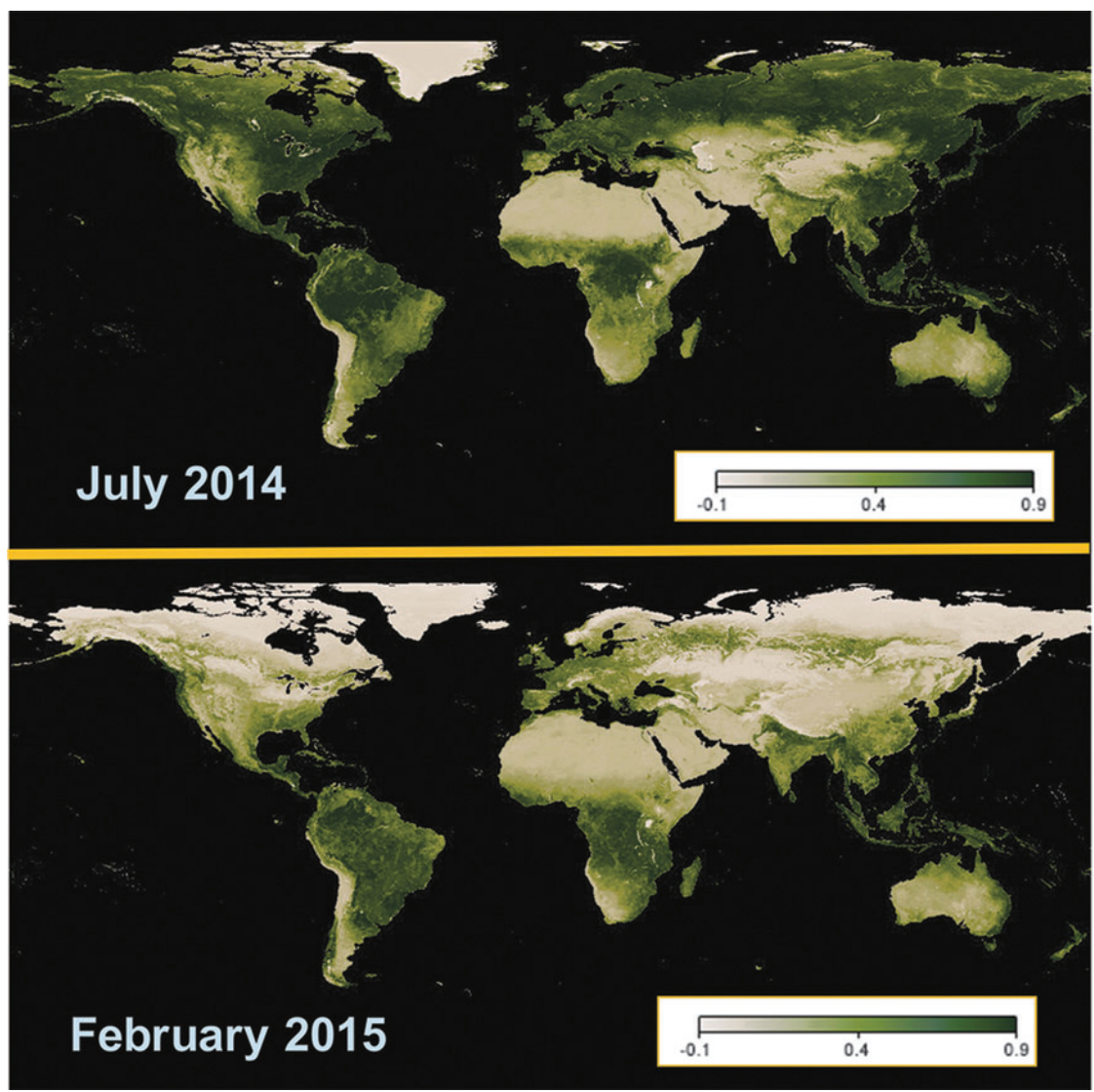

**FIG. 16.** Seasonal change in global NDVI. These vegetation maps were generated from data taken by the Moderate Resolution Imaging Spectroradiometer (MODIS) on board NASA's Terra satellite and converted to NDVI values. Significant seasonal variations in the NDVI are apparent between northern hemisphere summer (July 2014; top) and winter (February 2015; bottom). Image credit: Reto Stockli, NASA Earth Observatory Group, using data from the MODIS Land Science Team (http://neo.sci.gsfc.nasa.gov).

### 7.1. Chemical disequilibrium

It has long been recognized that the chemical composition of Earth's atmosphere is far from equilibrium. Much effort has been directed at studying this disequilibrium to understand the biological inputs that can allow for, for example, the simultaneous existence of $CH_4$ and $O_2$. These two gases should rapidly oxidize to $CO_2$ and $H_2O$, so their persistent presence suggests a continual resupply of both gases from biological sources. This state of disequilibrium (sometimes termed "redox disequilibrium") has been proposed as a biosignature (Lederberg, 1965; Lovelock, 1965; Hitchcock and Lovelock, 1967; Sagan *et al.*, 1993), and is a chemical signature that can be used to guide the search for life in exoplanet atmospheres. Much effort has been directed at quantifying the degree of disequilibrium to try to disentangle abiotic (*e.g.*, photochemistry) and biotic (methanogenesis and OP) inputs. However, Simoncini *et al.* (2013) found that rather than quantifying the distance from disequilibrium, a useful parameter to examine is the power required to drive disequilibrium. If the residence time of the gases in the atmosphere is long, a relatively small amount of energy is required to establish a high degree of disequilibrium. Conversely, if the residence time of the gases is short, and the disequilibrium high, then a high degree of power, likely biological, is required to maintain that atmospheric composition.

The $O_2$-$CH_4$ (and the related $O_3$-$CH_4$) disequilibrium pair is the most highly cited and explored biosignature couple in Earth's atmosphere (*e.g.*, Sagan *et al.*, 1993) because it requires large fluxes of $CH_4$ (as the less abundant gas) to maintain, fluxes that are believed to be incompatible with abiotic sources. However, it is important to note that $CH_4$ is both substantially less abundant and less spectrally detectable in Earth's modern atmosphere than $O_2$. In addition, these two gases absorb most strongly at disparate wavelengths—$O_2$'s strongest feature is in the optical at 0.76 μm, while $CH_4$'s strongest absorption features lie deep into the NIR and MIR wavelength regimen. These challenges and others pose potential obstacles for observing the $O_2$-$CH_4$ biosignature couple in exoplanets (see Section 8 below).

While references to chemical disequilibrium in Earth's atmosphere typically focus on the $CH_4$/$O_2$ disequilibrium pair, Krissansen-Totton *et al.* (2016a) recently found that the largest disequilibrium in the Earth system in terms of free energy is the coexistence of substantial quantities of free $O_2$ and $N_2$ in the atmosphere in contact with a liquid water ocean. In an equilibrium state, $N_2$ and $O_2$ would be converted by, for example, lightning into $NO_x$, in turn oxidized into $HNO_3$, which is rapidly washed out of the atmosphere to form the thermodynamically stable $NO_3^-$ in the ocean. The fact that substantial quantities of both $N_2$ and $O_2$ exist simultaneously within the atmosphere–hydrosphere system suggests that nonequilibrium (living) processes are acting to reprocess these gases. To verify the presence of this potential disequilibrium signature, $O_2$, $N_2$, and a liquid water ocean would need to be detected. This combination is hypothetically detectable with known methods, through the optical-NIR absorption features of $O_2$ (given in Section 4.2.1), ocean glint inferred from departures of Lambertian reflectance as the planet spectrum evolves with phase (*e.g.*, Robinson *et al.*, 2010, 2014), and detection of $N_2$ through $N_2$-$N_2$ collisional absorption near 4.1 μm (Schwieterman *et al.*, 2015b) or retrievals of atmospheric mass through Rayleigh scattering or estimates of atmospheric refraction that can constrain $N_2$ abundances.

### 7.2. Biomass estimation

Because of the abundance of water as an electron donor for photosynthesis and the availability of light on Earth, OP dominates primary productivity and it is currently responsible for gross fixation of 100–175 Pg of inorganic carbon per year (Field, 1998; Welp *et al.*, 2011). Currently, ocean and land primary productivity account each for approximately half the global total. Oxygenic phototrophs, such as cyanobacteria on the early Earth, possibly generated comparable biomass in shallow nearshore areas, continental settings, and possibly in the open ocean (*e.g.*, Garcia-Pichel *et al.*, 2003). Such extensive areal coverage may bode well for detection of pigmented surface biosignatures on an inhabited exoplanet. In contrast, anoxygenic phototrophs are usually limited by the environmental availability of electron donors (*e.g.*, $H_2$, $H_2S$, $Fe^{2+}$). Initial calculations suggest that these types of communities might be two to three orders of magnitude less productive (with less surface biomass generated) than modern oxygenic ones (Des Marais, 2000; Kharecha *et al.*, 2005), possibly making their detection on exoplanets more difficult.

Beyond surface biosignatures whose detectability depends on percent areal coverage on the continents, near shore areas, and ocean, Seager *et al.* (2013a) developed a model to assess the detection of biogenic gases in the atmosphere that is linked to biomass estimates based on thermodynamic calculations. This model is not meant to apply to all types of biosignatures but primarily to biogenic gases that are by-products of metabolisms that obtain energy from chemical potential gradients in the environment and to biochemicals that have specialized functions other than energy capture. In these cases, the amount of biomass correlates with amount of activity. A high cycling rate otherwise does not necessarily correlate with a high net flux or a high biomass stock. Therefore, the model does not apply to oxidizing gases resulting from biomass building processes in which a large stock of inactive biomass or fixed carbon may be stored. A primary example is the biogenic gas $O_2$ from OP, for which large organisms such as trees may have a large carbon stock in woody tissue, while active foliage biomass is relatively small. At the planetary scale, $O_2$ has an extremely low net flux, and its large atmospheric accumulation on Earth is the result of long-term storage of fixed carbon in ocean sediments and soils on land (Catling, 2014).

For the relevant biological processes mentioned above, this kind of biomass estimation model can establish first-order estimates for the plausible biogenicity of these gases, rather than adopting terrestrial fluxes. Importantly, this type of framework can be employed as a plausibility assessment of the utility of a potential biosignature gas for a specific planet–star combination, given the prediction of the biomass necessary to produce a detectable signature. In this scenario, the UV properties of the host star (which drives photochemistry) and the observing capabilities of the telescope are input parameters. The flux of a given biosignature gas (*e.g.*, $N_2O$, $CH_3Cl$) is increased until a detectability threshold is met. That flux is linked to a thermodynamic estimate of biomass, which can then be evaluated for plausibility. This was done for a range of scenarios presented in Seager *et al.* (2013a).

### 7.3. Applications of network theory to biosignatures

Another possibility for detecting life on an exoplanet is to look at the interactions between atmospheric constituents. By using techniques developed for graph theory and network analysis, the set of reactions present in an atmosphere can be converted into a chemical reaction network (CRN). In a CRN, chemical species are represented by nodes in a graph, with the reactions between them forming "edges," or connections between these nodes. The topological properties of the resulting network can then be measured and may include assortativity, which is the measurement of a node to attach to others that are similar in some way; average clustering coefficient, which indicates the degree that nodes in a graph tend to cluster together; and mean degree, or the average number of a degrees a node in the network has. In addition, other statistical properties, such as the degree distribution (how edges are distributed across the population of nodes, and how this distribution scales), can also provide insights into the structure of the network.

Preliminary studies by Solé and Munteanu (2004) examined the CRNs of every significant planetary atmosphere in the Solar system, and found that Earth's was unique: its topology displays a hierarchical and modular structure, whereas the CRNs of other planets appear more random. Further investigation by Holme *et al.* (2011) found similar results, suggesting that the degree, distribution, and scaling of Earth's reaction network are more like those of metabolic reaction networks than other planetary atmospheres. In addition, Solé and Munteanu (2004) reported that Earth's atmosphere is more modular, also more like biology.

Holme *et al.* (2011) further speculated that the differences between the network structure of Earth's atmosphere and others may be due to the comparatively small size of the data set for most planetary CRNs. Nonetheless, given that similar topological signatures have already been identified in biological networks (Jeong *et al.*, 2000)—a feature that in and of itself has been suggested as a potential biosignature (Jolley and Douglas, 2012)—and that the topology also appears to influence chemical disequilibrium in the atmosphere (Estrada, 2012), these findings suggest the possibility that the presence of a global biosphere influences the topology of a planet's atmospheric CRN in a significant, potentially detectable way.

While further study is required to validate this technique (especially in terms of comparing the topology of nonbiological systems, such as hot Jupiters) and establish its sensitivity, it holds tremendous promise as a tool for assessing exoplanet habitability and potential influence of a global biosphere. A following companion article contains further discussion of this topic (Walker *et al.*, 2018).

## 8. Cryptic Biospheres: "False Negatives" for Life?

While it can be argued that the evidence for life on modern Earth is strongly remotely detectable from the presence of abundant $O_2$, this will not necessarily be the case for every inhabited planet. Biospheres that persist only on chemosynthetic metabolisms will have orders of magnitude lower productivity than photosynthetic ones (Des Marais, 2000), and chemosynthesis reduces chemical disequilibrium rather than enhancing it (Seager *et al.*, 2012). Moreover, many known microorganisms and even entire ecosystems are cryptic to surface spectral detection, such as endolithic communities. Photosynthetic endoliths may live in fractures under rock surfaces, ruling out surface biosignatures, although they may emit $O_2$ if oxygenic (Cockell *et al.*, 2009a). Subsurface oceans in the icy moons of the outer Solar system may host redox-driven microbial biospheres cryptic to remote detection (*e.g.*, Lipps and Reiboldt, 2005) with confirmation only possible with *in situ* exploration.

Studies of the early Earth, a persistently habitable and inhabited planet, likewise sound a cautionary note for expectations of the general detectability of exoplanet biosignatures, particularly for the utility of the $O_2$-$CH_4$ biosignature couple (Reinhard *et al.*, 2017; see Fig. 17). The Archean Earth's atmosphere was significantly more reducing than today's and contained no detectable $O_2$ or $O_3$. While the Archean eon's (4.0–2.5 Ga) atmosphere likely contained significant detectable amounts of $CH_4$, it is unclear whether this alone would be an unequivocal sign of life (Arney *et al.*, 2016). The Proterozoic eon (0.5–2.4 Ga) has been viewed in recent studies as perhaps an archetypical model for a highly detectable disequilibrium biosphere, believed to have both detectable $O_2$ (10%) and much higher $CH_4$ ($\sim$100 ppm) than modern Earth (*e.g.*, Kaltenegger *et al.*, 2007). However, recent geochemical evidence based on chromium isotopes suggests that Proterozoic $O_2$ was quite low (perhaps <0.1% of PAL) (Planavsky *et al.*, 2014b; Lyons *et al.*, 2014), while biogeochemical modeling studies of sulfate cycling in the ancient ocean suggest that $CH_4$ was also much lower (1–10 ppm) (Olson *et al.*, 2016). If these estimates of low $O_2$ are correct, Fig. 17 shows that $O_3$ would have resided at the cusp of detectability during the mid-Proterozoic eon, and would have only been potentially detectable in the UV (Reinhard *et al.*, 2017). (Although the $O_3$ concentration is relatively insensitive to $O_2$ abundances over several orders of magnitude, this relationship breaks down at the lowest published limits for Proterozoic $O_2$; *i.e.*, $pO_2 < 0.1\%$ PAL).

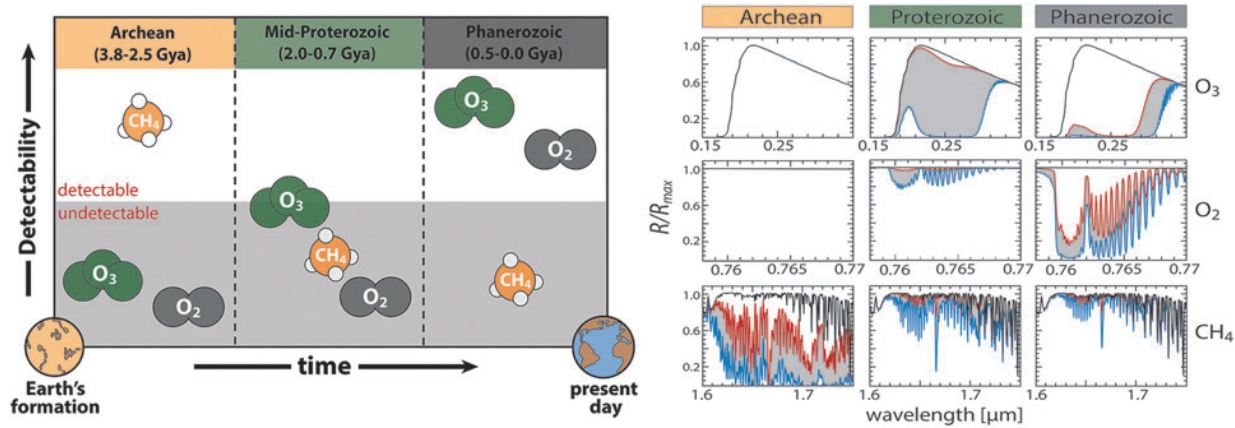

**FIG. 17.** Potential "false negatives" for remote life detection. Left: conceptual figure illustrating the difficulty of detecting the $O_2$-$CH_4$ disequilibrium signature through Earth history. Right: synthetic spectra, at 1 cm$^{-1}$ resolution, of the $O_3$ Huggins–Hartley band (0.25 μm), $O_2$-A band (0.76 μm), and 1.65 μm $CH_4$ band at different geologic times assuming gas abundances consistent with geochemical proxies and/or biogeochemical modeling constraints. Black represents no absorption by the indicated gas, red a lower limit, and blue an upper limit, given reasonable uncertainties in geochemical proxies. Adapted with permission from Reinhard *et al.* (2017).

As vascular land plants have only existed for the last ~470 million years (Kenrick and Crane, 1997), surface signatures of life on early Earth may have been significantly muted unless microbial mats or other microbial structures had significant areal extent comparable to, or larger than, vegetation today (*e.g.*, Sanromá *et al.*, 2013). There is evidence of microbial mats inhabiting land surfaces over 1 billion years ago (Horodyski and Knauth, 1994; Prave, 2002; Beraldi-Campesi, 2013). Investigators have been exploring surface biosignatures resulting from anoxygenic phototrophs and cyanobacteria growing in continental, marine marginal, and open ocean settings and are addressing the percent areal coverage necessary for detection through different atmospheric compositions and cloud coverage levels (Sanromá *et al.*, 2014; Parenteau *et al.*, 2015). However, the possibility exists that significant portions of Earth history, if not the entire window before the second rise of $O_2$ (~0.54 Ga), could represent a potential "false negative" for remotely detectable life, at least through $O_2$ and the $O_2$-$CH_4$ disequilibrium couple (Reinhard *et al.*, 2017) and perhaps for surface signatures as well. No matter how common or rare life is in the Universe, it is unlikely that every inhabited planet will have remotely detectable biosignatures (Cockell, 2014). This lesson from Earth history should be kept in mind when creating statistical frameworks for the prevalence of life given negative results and when estimating biosignature yields for surveys of candidate exoEarths (Stark *et al.*, 2014, 2016; Léger *et al.*, 2015). In addition, these results further reinforce the importance of broad spectral capabilities (UV-VIS-NIR-MIR) and high spectral resolving power of exoplanet characterizing missions and facilities, which would enhance the chance of characterizing biosignatures on worlds such as the early Earth.

### 9. Prospects for Detecting Exoplanet Biosignatures

The actual detectability of any of the biosignatures presented in this review will be related to a variety of factors, including those related to the observing telescope architecture (*e.g.*, aperture, coronagraph throughput, instrument sensitivity), target system distance and qualities (*e.g.*, spectral type, exozodiacal light), planetary parameters (*e.g.*, size, albedo, composition), and the accessible spectral range that will depend both on variables related to the telescope and target system. In general, the detection of exoplanet biosignatures is expected to follow a trajectory where interesting and accessible (via remote sensing) planet candidates are identified, for example, by the TESS mission (2018 launch; Ricker *et al.*, 2014; Sullivan *et al.*, 2015), the PLATO 2.0 mission (2026 launch; Rauer *et al.*, 2014), or ground-based radial velocity (RV) or transit surveys, followed by intensive characterization by the JWST mission (2020 launch, Deming *et al.*, 2009; Stevenson *et al.*, 2016), extremely large ground-based telescopes (ELTs; first light in mid-2020s), or a future direct-imaging mission such as the envisioned LUVOIR/HDST or HabEx class missions (~2030s) (Dalcanton *et al.*, 2015; Mennesson *et al.*, 2016).

In addition, ELTs and LUVOIR/HabEx would have the capability to both independently identify initial candidates and characterize them in search of biosignatures. JWST is likely to characterize only a handful of potentially habitable planets with the fidelity required to detect biosignature gases or signs of habitability (Cowan *et al.*, 2015). Importantly, transmission spectroscopy can only probe for atmospheric habitability markers and biosignatures at a single phase, excluding the detectability of all potential surface biosignatures and most proposed temporal biosignatures. With a planned 2025 launch, WFIRST could potentially characterize Earth-sized planets in the HZs of Alpha Centauri A and B, and possibly a handful of other nearby stars, if flown with a star shade and depending on the in-flight performance of instrumentation (Spergel *et al.*, 2015). The available wavelength range is an essential consideration for exoplanet biosignature assessment, because many gases will not create measurable spectral impacts in the UV-visible range. There is no space-based mission currently being studied that will possess the capability of characterizing the MIR spectrum of an Earth-sized planet, although the TPF-I concept was studied in the past (Beichman *et al.*, 2006; Lawson *et al.*, 2006). However, a combination of LUVOIR/HDST, which may cover a spectral range inclusive of 0.2–2.0 μm, and ground-based telescopes sensitive in the MIR, could cover most of the electromagnetic spectrum relevant for biosignatures.

A companion article (Fujii *et al.*, 2018) discusses upcoming missions and observatory capabilities in far more extensive detail, and provides a robust time line for expected biosignature detection capabilities. In addition, other companion articles (Catling *et al.*, 2018; Walker *et al.*, 2018) provide a foundation for robust frameworks for defining and evaluating biosignature detections.

### 10. Summary

We have provided an overview of potential signs of life on exoplanets, including gaseous, surface, and temporal biosignatures. The most detectable signs of life will likely result from a photosynthetic biosphere. Biosignature gases in Earth-like ($N_2$-$H_2O$-$CO_2$) atmospheres include $O_2$, $O_3$, $CH_4$, $C_2H_6$, $N_2O$, $CH_3Cl$, $CH_3SH$, DMS, and DMDS, although any individual gas alone is likely insufficient for biosignature confirmation due to potential false positive scenarios. Organic aerosols may be suggestive of life in atmospheres high in $CO_2$, while $NH_3$ may be a biosignature gas in $H_2$-dominated terrestrial atmospheres, providing false positives can be ruled out. Overlaps between absorbing wavelengths of key gases (*e.g.*, $O_3$ with $CO_2$ and $CH_3Cl$) and the potential abiotic production of certain gases caution against reliance on any single spectral feature, and indicate a wide spectral range is necessary for biosignature characterization. The interpretation of gaseous signatures will depend on the redox state of the atmosphere, which will determine which disequilibrium signatures are feasible.

The environmental context will also play a key role in interpreting potential gaseous biosignatures. For example, the $N_2$-$O_2$-ocean disequilibrium signature requires the detection of surface liquid water, perhaps through glint. The most well-studied surface biosignature continues to be the VRE. Detecting an exact analogue to Earth's disk-averaged VRE signature will likely require 1% spectrophotometric precision and ~10% or more cloud-free surface coverage of exovegetation. Analogues to the VRE, "edge" biosignatures, may be produced by photosynthetic or nonphotosynthetic pigments or structures and occur throughout the visible and NIR spectrum. Linear and circular polarization signatures and contextual information could be used to rule out false positives for surface biosignatures. Temporal biosignatures may include seasonal modulation in biologically mediated gases such as $CO_2$ or $O_2$, changes in surface signatures such as analogues to the VRE, or direct emission of light by organisms (*e.g.*, bioluminescence, fluorescence). In general, temporal biosignatures are less well studied than gaseous or surface biosignatures and additional work is necessary to elucidate the range of applicability for this category of signatures.

Earth history suggests that not all inhabited planets with global biospheres will necessarily have remotely detectable signs of life, and that broad spectral capabilities and high resolving power will increase the chances of success. Updates to the disequilibrium framework, new biomass estimation models, discoveries of greater diversity of photosystems in OP, and network theory applied to atmospheric biosignatures offer novel or evolving frameworks, which may enhance our ability to characterize and correctly interpret exoplanet biosignatures in the future.

## Acknowledgments

The authors thank the NASA Astrobiology Program and the Nexus for Exoplanet System Science (NExSS) for their support of the NExSS Exoplanet Biosignatures Workshop. Conversations at this workshop, held in the summer of 2016 in Seattle, formed the basis for the drafting of the five review articles in this issue. They also thank Mary Voytek, the senior scientist of NASA Astrobiology, for her leadership of NExSS and her feedback on our organization of the workshop and article. E.W.S. is additionally grateful for support from the NASA Postdoctoral Program, administered by the Universities Space Research Association. This work was also supported by the NASA Astrobiology Institute, including the VPL under Cooperative Agreement Number NNA13AA93A and the Alternative Earths team under Cooperative Agreement Number NNA15BB03A. S.D. acknowledges support from NASA exobiology grant NNX15AM07G. The research of R.H. was carried out at the Jet Propulsion Laboratory, California Institute of Technology, under a contract with the National Aeronautics and Space Administration. The authors thank the following individuals for helpful comments, suggestions, and discussion during the community review period: Tanai Cardona, Anthony Del Genio, Stephen Kane, and Enric Pallé. Finally, the authors are thankful for the helpful comments from two anonymous referees, which allowed us to further improve the paper.

## Author Disclosure Statement

No competing financial interests exist.

Address correspondence to:
*Edward W. Schwieterman*
*Department of Earth Sciences*
*University of California*
*Riverside, CA 92521*

*E-mail:* eschwiet@ucr.edu

**Abbreviations Used**

1D = one-dimensional
3D = three-dimensional
$C_2H_6$ = ethane
$CH_3$ = methyl
$CH_3SH$ = methanethiol
$CH_4$ = methane
CRN = chemical reaction network
DMDS = dimethyl disulfide
DMS = dimethyl sulfide
ELTs = extremely large ground-based telescopes
GCM = General Circulation Model
GOE = Great Oxidation Event
$H_2S$ = hydrogen sulfide
$HNO_3$ = nitric acid
HZ = habitable zone
JWST = James Webb Space Telescope
MIR = midinfrared
$N_2O$ = nitrous oxide
NDVI = Normalized Difference Vegetation Index
NExSS = Nexus for Exoplanet System Science
$NO_3^-$ = nitrate
$NO_X$ = nitrogen oxides
NUV = near-ultraviolet
OH = hydroxyl
OP = oxygenic photosynthesis
PAL = present atmospheric level
PSI = photosystem I
PSII = photosystem II
RC = radiative–convective
SETI = Search for Extraterrestrial Intelligence
$SO_2$ = sulfur dioxide
TPF-I = Terrestrial Planet Finder—Infrared mission
UV = ultraviolet
UV-VIS-NIR-MIR = ultraviolet, visible, near-infrared, and midinfrared
VPL = Virtual Planetary Laboratory
VRE = vegetation red edge